\documentclass[letterpaper, 10 pt, conference, onecolumn]{ieeeconf}  

\IEEEoverridecommandlockouts                              
\usepackage{graphicx}
\usepackage{float}
\usepackage{multirow}
\usepackage{algorithm}
\usepackage{algorithmic}
\usepackage{lscape}
\usepackage{listings}
\usepackage[table,xcdraw]{xcolor}
\usepackage{pbox}

\lstdefinestyle{DOS}
{
    backgroundcolor=\color{white},
    basicstyle=\scriptsize\color{black}\ttfamily
}

\title{\LARGE \bf
Improving web element localization by using a large language model
}

\author{Michel Nass$^{1}$, Emil Al\'egroth$^{2}$ and Robert Feldt$^{3}$
\thanks{$^{1}$H. SERL, Blekinge Institute of Technology
        {\tt\small michel.nass@bth.se}}%
\thanks{$^{2}$P. SERL, Blekinge Institute of Technology
        {\tt\small emil.alegroth@bth.se}}%
\thanks{$^{3}$P. SERL, Blekinge Institute of Technology and Chalmers University of Technology
        {\tt\small robert.feldt@chalmers.se}}%
}

\begin{document}

\maketitle
\thispagestyle{empty}
\pagestyle{empty}

\begin{abstract}
\textbf{Context:}
Web-based test automation heavily relies on accurately finding web elements.
Traditional methods compare attributes but don’t grasp the context and meaning of elements and words.
The emergence of Large Language Models (LLMs) like GPT-4, which can show human-like reasoning abilities on some tasks, offers new opportunities for software engineering and web element localization.

\textbf{Objective:}
This paper introduces and evaluates VON Similo LLM, an enhanced web element localization approach.
Using an LLM, it selects the most likely web element from the top-ranked ones identified by the existing VON Similo method, ideally aiming to get closer to human-like selection accuracy.

\textbf{Method:}
An experimental study was conducted using 804 web element pairs from 48 real-world web applications.
We measured the number of correctly identified elements as well as the execution times, comparing the effectiveness and efficiency of VON Similo LLM against the baseline algorithm.
In addition, motivations from the LLM were recorded and analyzed for all instances where the original approach failed to find the right web element.

\textbf{Results:}
VON Similo LLM demonstrated improved performance, reducing failed localizations from 70 to 40 (out of 804), a 43\% reduction.
Despite its slower execution time and additional costs of using the GPT-4 model, the LLM’s human-like reasoning showed promise in enhancing web element localization.

\textbf{Conclusion:}
LLM technology can enhance web element identification in GUI test automation, reducing false positives and potentially lowering maintenance costs.
However, further research is necessary to fully understand LLMs’ capabilities, limitations, and practical use in GUI testing.
\end{abstract}

\begin{keywords}
GUI Testing, Test Automation, Test Case Robustness, Web Element Locators, Large Language Models
\end{keywords}

\section{Introduction}

Software testing plays a vital role in ensuring the quality of software applications.
However, testing is often a time-consuming and expensive process in practice~\cite{grechanik2009maintaining,grechanik2009creating}.
By leveraging automation, organizations can run tests more frequently, improve test coverage, and thereby identify more defects faster, with positive impacts on software lead times and software quality~\cite{olan2003unit,adamoli2011automated,alegroth2013transitioning}.

Automation is applied in various types of testing, but one of its primary uses in practice is in automated regression testing.
Regression testing allows testers to evaluate the quality of each software release.
Typically, at higher levels of system abstraction, such as the Graphical User Interface (GUI) level, testers create a suite of test scripts that simulate end-user scenarios and verify the application under test's (AUT) correct behavior by using automated oracles~\cite{liebel2013state,mahmud2014design}.
However, it is common for new software releases to introduce changes that can break existing automated regression tests, which require maintenance efforts and costs to update and repair the test scripts.
The maintenance cost is exceptionally high when testing an application through its GUI, as GUIs frequently change between releases~\cite{tonella2014recent,alegroth2017long,dobslaw2019estimating}.
In addition, GUI scripts are subject to breaking from changes to the underlying logic and architecture of the AUT that modifies its behavior.

Furthermore, GUIs are primarily designed for human interaction (i.e., not machine-to-machine communication), which presents additional challenges for automation, such as synchronization between the test scripts and the AUT.
These challenges, although present, are not considered as prominent in lower-level testing techniques like unit testing~\cite{olan2003unit}.

Test script robustness is one of the most reported challenges in web test automation~\cite{nass2021many}.
The challenge involves making tests resilient to smaller changes to the AUT that should not affect the test execution while still allowing the tests to detect significant differences that could potentially be defects.
Many solutions that increase the robustness of locating web elements have been proposed for mitigating this challenge~\cite{choudhary2011water, montoto2011automated, leotta2014reducing, leotta2016robula+, thummalapenta2012automating, leotta2015using}.
Some of the more recent approaches use similarity scores to identify the most similar web element to a target.
This is done by using previously stored properties (i.e., extracted from the corresponding web element in a previous version of the web application) and comparing the stored properties to the updated web elements~\cite{nass2022similarity, nass2023robust}.
The web element with the highest score is assumed to be the most likely web element to use in an interaction (e.g., a click or type action).
While conventional algorithms (i.e., non-AI) can be used for finding similarities between web elements, they still typically lack knowledge about how web applications work and the semantic meaning of texts (i.e., skills possessed by a human tester).
Being able to tell if different words or sentences have the same meaning or that two different web elements have contextual similarities (e.g., are closely located or are interchangeable solutions) could be a powerful feature in a testing tool.
For example, assume a button in a web interface that changes the caption from 'Submit' to 'Send' in an updated version.
A script that relies on the button caption to identify the next action would likely not find the new caption identical to the old caption without some form of semantic understanding, causing a false positive (i.e., a failed script execution).
On the other hand, if a test tool could reason that the captions still have the same meaning (i.e., in that specific context), they could perceivably carry on without failing the test execution.

Large language models (LLMs) are trained on vast amounts of data and utilize deep learning techniques to capture linguistic patterns and dependencies \cite{vaswani2017attention}.
We have only begun to explore the possibilities of using LLMs in test automation.
One such example is SocraTest, a vision of a framework for conversational testing agents that could aid a human software tester by performing tasks autonomously \cite{feldt2023towards}.
Recent studies utilize natural language processing (NLP) with heuristic search and the DOM structure to identify web elements in web applications \cite{kirinuki2022web} or use LLMs to generate text inputs for GUI applications based on semantic understanding and GUI application context \cite{liu2022fill, liu2023chatting, wu2023screen}.
The proposed solution in this paper is based on the hypothesis that we can improve web element localization even further by combining an LLM with a traditional algorithm to take advantage of some of the benefits of the LLM, e.g., its assumed semantic understanding and contextual awareness, while utilizing the speed of the conventional algorithm.

The specific contributions of this paper are:

\begin{itemize}
\item A novel approach that can improve web element localization by utilizing a large language model.
\item An empirical study that shows the effectiveness and efficiency of the proposed approach compared to the baseline approach.
\item A qualitative content analysis on the motivations gathered from the LLM, explaining the main aspects used when comparing the similarity of two web elements.
\end{itemize}

This paper is structured as follows.
Section~\ref{LLMs} gives a short introduction to large language models.
Section~\ref{Similo algorithm} covers the details of both previous versions and the proposed enhancement to the Similo algorithm.
The design, research questions, and procedure of the empirical study are presented in Section~\ref{Methodology}, and the results in Section~\ref{Results}.
We then discuss results in Section~\ref{Discussion} and state conclusions and future work in Section~\ref{Conclusions}.
Section~\ref{Related Work} presents related work.

A package for replicating the experiment is available for download from~\cite{reppack}.

\section{Large Language Models}\label{LLMs}

Large language models (LLMs) like GPT-4 have revolutionized Natural Language Processing (NLP) by leveraging the Transformer architecture \cite{vaswani2017attention}.
This groundbreaking approach replaced traditional recurrent neural networks (RNNs) with a self-attention mechanism, enabling the models to capture long-range dependencies efficiently.
These models are pre-trained on vast amounts of data, allowing them to grasp the meaning of input prompts and generate text.
Notable examples of recent LLMs include OpenAI's Generative Pre-trained Transformer (GPT-3, GPT-3.5, and GPT-4)~\cite{bubeck2023sparks} and Google's Pathways Language Model (PaLM)~\cite{chowdhery2022palm}.
ChatGPT is a sibling model to the InstructGPT model, which is an improved version of GPT-3 that has been fine-tuned and trained with human feedback~\cite{christiano2017deep} to improve its ability to follow instructions~\cite{ouyang2022training}.
We can also use ChatGPT as an interface to the newer GPT-4 model, which is several magnitudes larger (i.e., about 1000 times) than previous GPT models and performs close to human-level on some tasks~\cite{bubeck2023sparks}.
We have included a more detailed comparison between the two latest versions of GPT in Section \ref{Selecting Model}.

\section{Similo}\label{Similo algorithm}

VON Similo is a web element localization algorithm that uses a multi-locator approach, similar to previous works, e.g., Leotta et al.~\cite{leotta2016robula+}.
In contrast to single-locator solutions, multi-locators use multiple properties of a web element, such as ID, XPath, label, and tag, to find a target.
This is achieved by comparing the properties of each candidate web element on a web-page with the desired properties of a target element (i.e., the correct candidate), resulting in a similarity score. 
A heuristic is then applied that the web element with the highest similarity is the most likely candidate to be a match.

VON Similo also utilizes the concept of visually overlapping nodes (VON), which makes use of the hierarchical structure of web elements in modern web applications and their representation in a document object model (DOM)~\cite{wood1998document}. 
The VON concept considers that multiple DOM nodes (i.e., web elements) are often visually overlapping (i.e., displayed in the same visual area in the web browser) and conjointly represent the same visual web element to the user.
These conjoint elements share or have similar properties, e.g., overlapping areas, coordinates, and similar XPaths.
This implies that interactions (e.g., a click) on the area represented by any of these overlapping nodes will yield the same GUI state transition (i.e., event).
As such, any of the nodes can be used to execute an automated test case, effectively increasing the number of valid web elements for an interaction from a single element to the number of overlapping elements in a visual area.
This increase in targets improves the probability of finding a web element after changes to the tested application, thereby increasing the test execution robustness.

In this paper, we use the following nomenclature:

\begin{itemize}
\item \textbf{Properties}: attributes and other information (e.g., location, size, XPath, etc.) that can be extracted from a web element.
\item \textbf{Candidate}: a web element containing properties that can be evaluated by VON Similo. Candidates are typically captured from the currently active (i.e., visible) web page.
\item \textbf{Desired properties}: the properties we are looking for in a candidate. The desired properties are often captured or recorded from a target in a previous version of the SUT (i.e., when the test script was created or maintained).
\item \textbf{Similarity score}: a score representing the distance in similarity between two sets of properties, where a higher score represents higher similarity between two web elements.
\item \textbf{Visual web element}: one or many DOM nodes that overlap visually, according to the Visual Overlap heuristics defined by the VON-Similo algorithm (described in Section \ref{von_approach}).
\end{itemize}

\subsection{Standard Similo}

\begin{figure}[H]
  \centering
  \includegraphics[width=0.8\textwidth]{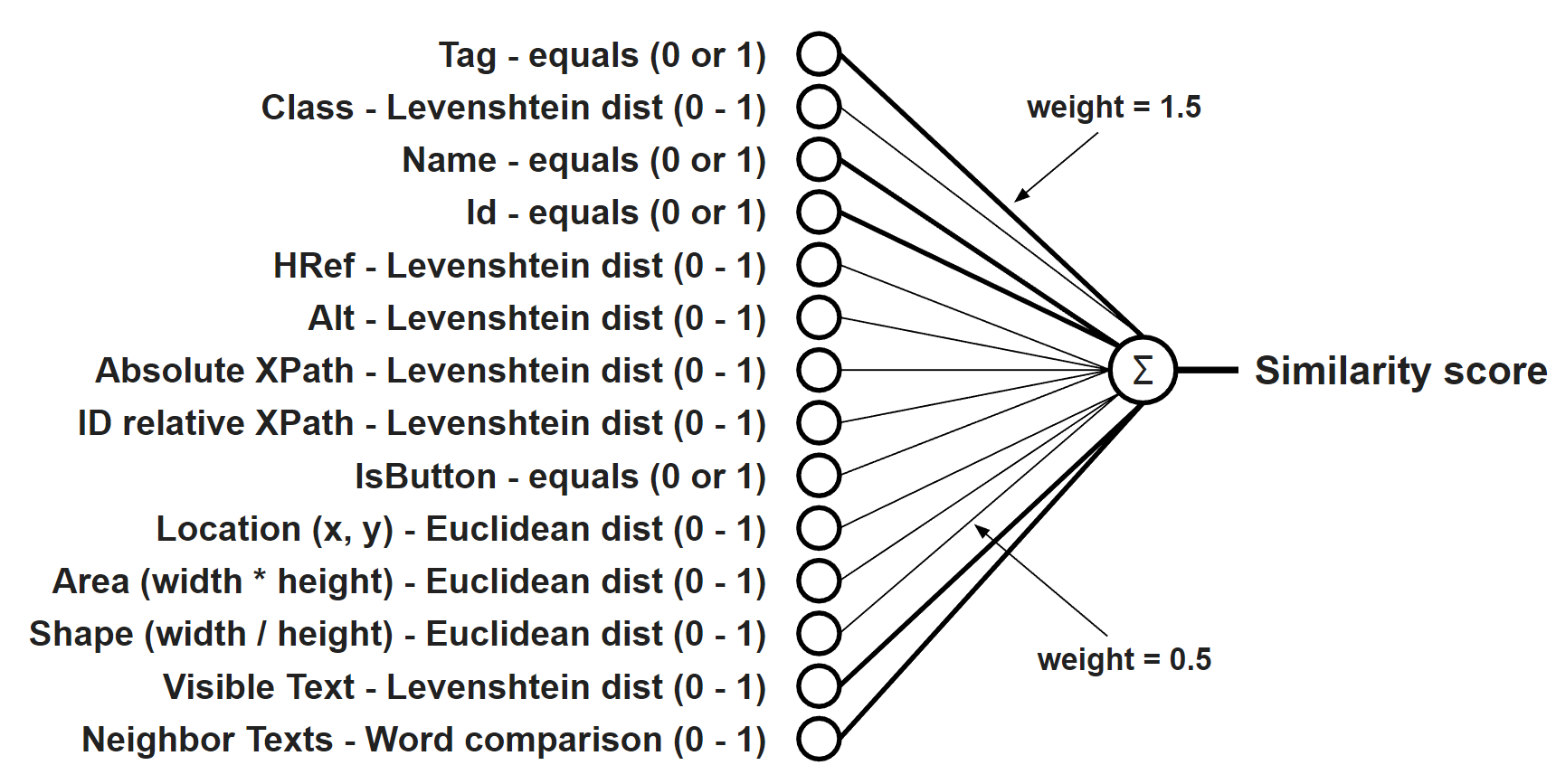}
  \caption{Graphical representation of the computation of similarity score between two different sets of web element properties.}
  \label{similo_original}
\end{figure}

VON Similo is based on the initial version of the Similo multi-locator algorithm proposed by Nass et al. \cite{nass2023similarity}.
Similo attempts to identify the web element among a set of candidates that is most similar to the desired properties.
The desired properties are often gathered or recorded from a previous release of the same AUT but can be any set of properties.
Candidate web elements are typically retrieved from the current (i.e., visible) web page.
The standard version of Similo used 14 properties, listed in Figure \ref{similo_original}.
Each property is associated with a comparison operator and a weight (also included in Figure \ref{similo_original}).
The comparison operator compares the property value of a candidate with the desired property value and returns an output value \textit{between} zero and one (or binary zero \textit{or} one).
Using the output values, a similarity score is then calculated for each candidate by summarizing the weight multiplied by the result from the comparison operator for all 14 properties.
After comparison of all candidate element scores, Similo then returns the candidate with the highest similarity score, assumed to be the most similar web element to the target element with the desired properties.
Optionally the algorithm can output a ranked list of candidates from higher to lower similarity scores.

\subsection{VON Similo} \label{von_approach}

\begin{figure}[H]
  \centering
  \includegraphics[width=0.8\textwidth]{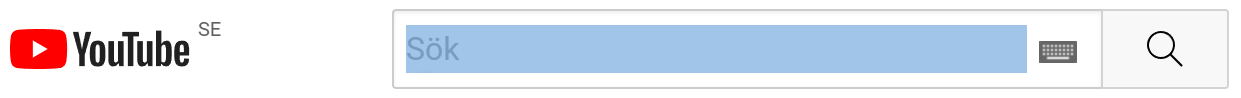}
  \caption{The YouTube search bar.}
  \label{youtube_search_bar}
\end{figure}

The concept of visually overlapping nodes (VON) can be applied to Similo to increase the likelihood of locating the correct web element (i.e., according to the oracle) \cite{10132199}.
To illustrate the VON concept with an example, Figure \ref{youtube_search_bar} contains a picture of the search bar on YouTube.

The light-blue area of the image contains two DOM elements in a hierarchy.
A simplified version of that DOM structure is shown in Listing~\ref{listing:dom}.

\begin{lstlisting}[style=DOS, breaklines=true, numbers=none, caption={DOM hierarchy of the YouTube search bar.}, label=listing:dom]
<div class="sbib_b" id="sb_ifc50">
    <input id="search" name="search_query">
</div>
\end{lstlisting}

As can be seen from this example, what visually appears to be only one element is actually represented by a div element containing an input element (i.e., two DOM elements in a hierarchy).
This exemplifies how modern web pages are structured and presents a problem when selecting an oracle that represents the correctly located DOM element (i.e., web element) since there is more than one to choose from.
The VON concept handles this problem by treating both of the DOM elements as equally correct by merging the properties of both elements together into one visual web element (i.e., a new virtual element).
Listing~\ref{listing:visual_web_element} illustrates how such a visual web element could be represented where the double pipe (i.e., ``OR'' operator) denotes that an attribute could have more than one value.

\begin{lstlisting}[style=DOS, breaklines=true, numbers=none, caption={Example representation of a visual web element.}, label=listing:visual_web_element]
<div || input class="sbib_b" id="sb_ifc50 || search" name="search_query"/>
\end{lstlisting}

There are two benefits to the VON approach.
First, it reduces the number of candidate web elements (i.e., since there are typically fewer visual web elements than DOM elements on a web page), resulting in a higher probability of locating the correct one.
Secondly, merging the properties of all the DOM elements belonging to the same visual web element will result in a higher (or the same) similarity score than distributing the score on several DOM elements in the hierarchy (i.e., any contributions to the similarity score, when comparing the properties, is concentrated on the same visual web element instead of being distributed over several DOM elements).

Two web elements (W1 and W2) are considered to belong to the same visual web element if the following conditions are satisfied:

\begin{enumerate}
\item The ratio between the overlapping areas of the web elements on the web page, and the union of the areas of the two web elements, is higher than a set threshold value (0.85 was selected by Nass et al. \cite{10132199}). The ratio can be computed as: 

\[ \frac{\cap(R_1, R_2)}{\cup(R_1, R_2)} \]

where: $R_1$ and $R_2$ are the rectanglular areas---Calculated using the coordinates (i.e. x and y) and size (i.e. width and height)---where the elements are visible on the web page.
The intersection of these areas thereby represents the size (in pixels) of the common area occupied by $R_1$ and $R_2$, and the union represents the size (in pixels) of the total area occupied by $R_1$ and $R_2$.

\item The center of the web element W2 is contained in the rectangle R1.
Note that this condition is always true if the threshold is greater than 0.5.
\end{enumerate} 

VON Similo (i.e., the VON concept applied on Similo) uses the same set of properties as in Figure \ref{similo_original} with the difference that each property value can take multiple values instead of just one, as in the Similo case.
A property that holds more than one value is compared several times (i.e., one time per value).
Assuming we would like to compare the Tag property in the web elements W1 and W2, we would need to perform N*M comparisons assuming that the Tag property in W1 contains N values and the Tag property in W2 contains M values.
The highest (i.e., best) comparison outcome of the N*M comparisons is selected as the result and appended to the similarity score.
As such, the final score can be comprised of the comparator outcomes of property values from multiple DOM elements joined in the new virtual element.

For example, assume that the Tag property values are 'div' and 'span' for W1 and that the corresponding property values are 'span' and 'button' for W2.
Comparing all the combinations (i.e., four) will result in a match (i.e., 'span') and return the value one from the equals comparison operator (See Figure \ref{similo_original}).

\subsection{Limitations of Similo and VON Similo}

While Similo and VON Similo increase the tolerance to changes (i.e., robustness), there are still situations where the algorithms fail to find the web element specified by the human oracle.
Our hypothesis is that humans possess reasoning capabilities, e.g. semantic, logical or contextual, about language and web applications that the algorithms lack.
For example, assume that a button changed the caption (i.e., visible text) from 'Save' to 'Store'.
A human would likely consider them to be equivalent buttons since the semantic meaning (i.e., purpose) is still the same, while the algorithm would struggle since the calculated distance between the two captions, e.g. using Levenshtein distance, would be quite large, negatively impacting the similarity score.
Another example is when a button changes from \{tag: 'input', type: 'button'\} to \{tag: 'button'\}.
If the tags were compared using the equals comparator, or even a distance comparator, the algorithm would not spot any similarities, while a context-aware human might know that a 'button' is a common replacement for an input field of type 'button' (i.e., an older standard).
The core hypothesis of this work is thereby that large language models (e.g., GPT-4), trained on a vast amount of texts and websites, possess some form of reasoning, akin to humans, which can complement conventional algorithms to improve their robustness.

\section{VON Similo LLM}

\begin{figure}[H]
  \centering
  \includegraphics[width=0.8\textwidth]{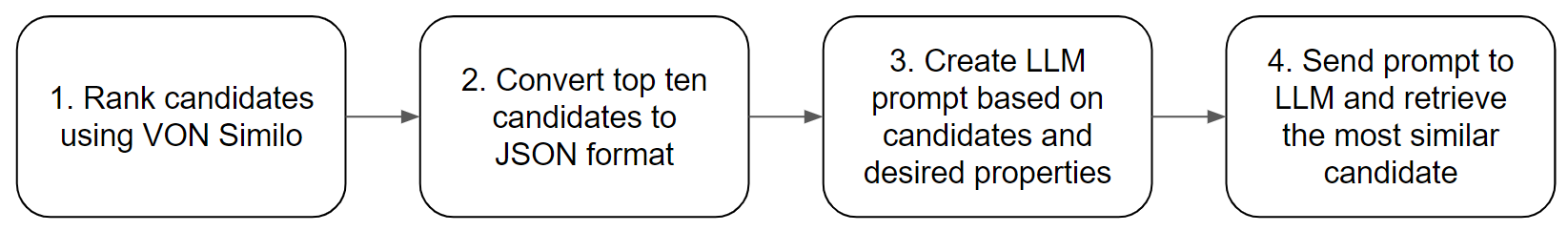}
  \caption{The VON Similo LLM process.}
  \label{steps}
\end{figure}

VON Similo LLM is an attempt to take advantage of the speed and determinism of a conventional algorithm, VON Similo, but improved by the language understanding\slash processing and assumed reasoning capabilities of a large language model (LLM).
In VON Similo LLM, we begin by ranking all the candidates present on the current web page with VON Similo.
This is done by comparing each element's properties to the desired properties (i.e., properties stored when creating or maintaining the test) of the target element.
Next, we extract the top ten candidates from the ranked list of candidates provided by VON Similo.
Each candidate in the top ten list and the desired properties of the target are then converted into a suitable format (i.e., we used JSON in the experiment since that should be a format familiar to an LLM).
A prompt is then generated for the LLM (i.e., GPT-4 in our case) containing instructions for the comparison, the ten candidates, and the desired properties of the target.
Figure \ref{steps} contains all the steps further detailed below.

\begin{enumerate}
\item
The first step in the process is to extract all the candidate web elements from the currently visible web page and rank them, based on similarity, using VON Similo.
VON Similo compares the desired properties with the properties of each of the candidates and produces a similarity score, as shown in Figure \ref{similo_original}.
The candidates are now sorted on similarity score, and the top ten continue to the next step.
We decided to limit the number of candidates to ten for our experiments to prevent the prompt from exceeding the usage quota and also reduce the runtime cost of utilizing the LLM API.
A usage quota is a limit of tokens spent over some time.
Quotas prevent a user of GPT-4 from accidentally paying too much money on prompts.

\item
The next step is to convert the ten candidates into a format that the LLM should be familiar with since that enables us to create a prompt without explaining the format.
We decided to use JSON since that is a commonly used format when communicating over the Internet.
Instead of creating an array, we decided to place each JSON structure on a separate line.
Listing~\ref{listing:json} shows an example of a prompt containing ten candidates encoded in JSON format.

\item
The third step is to create a prompt that contains instructions on what we expect the LLM to do and what we would like as output.
Listing~\ref{listing:jsonstructure1} shows the prompt structure we used.
The first eleven rows of the prompt contain one line of instruction and ten lines of candidates in JSON format.
We also provide a unique widget id (i.e., incremental count) with each candidate to simplify the output.
Next, we add the instruction that we expect the LLM to return with the widget id to the candidate most similar to the desired properties, also converted into JSON format.
Listing~\ref{listing:jsonstructure2} shows an alternate prompt structure used when asking the LLM to provide us with motivations explaining why this candidate is considered the most similar.
We have included a more complete example of the second prompt version in Listing~\ref{listing:json}.

\item
The final step is to send the prompt to the LLM.
The widget id of the most similar candidate and, optionally, the motivation of the choice, depending on the prompt used, are retrieved as output.
\end{enumerate} 

\begin{lstlisting}[style=DOS, breaklines=true, numbers=none, caption={Prompt structure used in experiment when asking for widget id only.}, label=listing:jsonstructure1]
Given the following candidate web elements (|| means that an attribute can have multiple values):
<10 candidates in JSON format>

find the one that is most similar to the element:
<desired properties in JSON format>
Answer with the widget_id number(digits) only, no explanation or text characters
\end{lstlisting}

\begin{lstlisting}[style=DOS, breaklines=true, numbers=none, caption={Prompt structure used in experiment when asking for widget id and motivations.}, label=listing:jsonstructure2]
Given the following candidate web elements (|| means that an attribute can have multiple values):
<10 candidates in JSON format>

find the one that is most similar (answer with the widget_id of the most similar and motivate why using a list) to the element:
<desired properties in JSON format>
\end{lstlisting}

\section{Methodology}\label{Methodology}
This section presents the research design, the research questions, and the research procedure of the empirical study performed to evaluate the benefits and drawbacks of VON Similo LLM compared to VON Similo in terms of effectiveness and efficiency.

The first objective of the experiment is to evaluate the difference in effectiveness between VON Similo LLM and the VON Similo approaches (i.e. when finding web elements in two different releases of the same web application).
The second objective is to compare the runtime performance (i.e., efficiency) of using the two approaches.
Finally, the third objective is to evaluate the motivations returned from the LLM to explain why the LLM found the chosen candidate element to be the most similar match to the correct candfidate.

\subsection{Research Questions}

The study aims to answer the following research questions:
\begin{itemize}
\item \textbf{RQ1:} What is the effectiveness of VON Similo LLM compared to the VON Similo approach in terms of finding correct web elements?
\item \textbf{RQ2:} What is the efficiency, measured as execution time, of VON Similo LLM compared to the VON Similo approach?
\item \textbf{RQ3:} What main aspects does a large language model use to improve web element identification when the conventional approach fails?
\end{itemize}

The first research question (RQ1) was answered by running both approaches on a set of 804 web element pairs extracted from old and new versions of 48 real-world web applications.
With a new version, we refer to a later iteration of a particular web application that has been subject to changes to its code or visual appearance that differentiates it from the older version (Further described in Section \ref{Extracting}).
Our hypothesis is that VON Similo LLM, using its reasoning capabilities, e.g., of semantic equivalence, logical patterns, or contextual information, would be able to correctly identify more correct candidates than VON Similo.

Next, research question 2 (RQ2) was answered by measuring the execution times of both approaches to determine the best matching web element.
We measured the execution time as the time taken from calling an approach (i.e., by providing it with the desired and candidate properties) and returning the most similar candidate.
Our hypothesis was that VON Similo would outperform VON Similo LLM in this aspect since VON Similo LLM utilizes the GPT-4 API (selected in Section~\ref{Methodology}), which, at the time of conducting the experiment, is relatively slow and restricted (i.e., in terms of requests per minute).
In addition to the actual overhead cost, this metric is assumed to give insights to allow us to discuss the current technology's industrial applicability.

Finally, we answered research question 3 (RQ3) by conducting a qualitative content analysis of the motivations gathered from the LLM, which aims to explain why the LLM found one candidate to be more similar to the correct candidate with some desired properties in more complex cases, i.e., in cases where the conventional approach (i.e., non-AI) failed to locate the correct candidate.
This analysis was restricted to cases where the conventional solution failed out of cost constraints, i.e., the output from the GPT-4 API is associated with a monetary cost per output token.

\subsection{Selecting Web Applications and Extracting Properties}\label{Extracting}

The web applications chosen for this experiment are the same 50 websites used by Nass et al. to evaluate previous versions of Similo~\cite{nass2022similarity}, taken from the Alexa top 50 list.
One of the applications from the top 50 list was deemed inappropriate due to its adult content, and one was a duplicate (i.e., two URLs pointing to the same web application), resulting in a final set of 48 web applications.
Additionally, we used the same web application versions, a new one and one 12 to 60 months older, as in the previous study~\cite{nass2022similarity}, accessed through the Internet Archive website \footnote{http://web.archive.org}. 
A scraping tool (developed in Java by the authors) was then applied to extract properties from all pairs of web elements that were perceived to be equivalent and available in both the old and new versions of each application.
These elements were chosen manually through inspection of the applications and then used as oracles for the study.
We manually included web elements for which the following criteria are met: (1) it is possible to perform an action on the web element, (2) the element can be used for assertions or synchronization by an automated testing tool, (3) the element belongs to the core features of the AUT, and (4) the element is present in both versions of the AUT's homepage (i.e., the page that the main URL points at).
Criteria (4) was necessary since the Internet Archive only stores static pages, meaning that javascript, databases, etc., do not always work.
Because the pages are static, they often, unintentionally, have diverse behaviors to newer versions of the AUTs.
Whilst this design choice of only using the homepage may delimit the generalizability of the results, we perceive this to be a minor threat since most homepages contain the same elements as other pages of an AUT.

Furthermore, this selection process implies that if a human could identify the web element in both versions of the web application, it was likely included.
This further implies that some web elements, which had been changed beyond recognition but which were still available, may have been overlooked during sampling.
However, due to the size of the sample set and the efforts spent to capture all pairs in the extraction process, we find this threat to be negligible.

We wish to highlight that the experiment only concerns the web element finding ability of the approaches.
We were not concerned with the types of interactions that can be performed on the elements nor how to utilize them for synchronization.
This objective further justifies our delimitation of only choosing web elements on the homepage of each website since we perceive these to be representative of other pages as well.
A perception that may not hold true for actions.

\subsection{Applying the VON Concept on the Extracted Properties}
In the next step of the research procedure, we applied the VON concept, described in Section~\ref{von_approach}, on each of the 804 web element pairs to add more values (i.e., from overlapping elements) to the target web element properties.
Due to the VON concept, property values of visually overlapping web elements will be merged (i.e., using an ``OR'' operation) if the ratio between the intersection and the union of the areas exceeds the threshold value (i.e., 0.85 in our case).
After applying the VON concept, many properties will contain several values (i.e., options) instead of just one, as when using the standard Similo approach.

\subsection{Selecting the Large Language Model} \label{Selecting Model}

Large Language Models (LLMs) are evolving quickly, and new versions are frequently released.
For our experiment, we decided that effectiveness (i.e., in identifying the correct candidate) was the most important aspect to evaluate (i.e., before efficiency and cost) since we expect the performance, availability (i.e., allowed requests per minute), and price to change in time as the services mature.
This design choice has a direct impact on RQ2, but we still perceive the results as valuable to get a snapshot of the currently available technology.
We also expect the effectiveness of LLMs to improve, but evaluating the effectiveness today will still provide us with a baseline for the future.
Therefore, we decided to select the most powerful LLM, in terms of effectiveness, available regardless of its efficiency, monetary cost (within reason), and limitations in requests per minute.
We also decided to go for an LLM provided by OpenAI due to its reputation and ease of access.
Table~\ref{llms} contains a comparison between the different versions currently provided by OpenAI (in April 2023, when we initiated the experiment).
As seen from Table~\ref{llms}, GPT-3.5-turbo is better in all aspects (e.g., cheaper and more requests allowed per minute), except for max tokens (4K vs. 8K for GPT-4).
The model size of GPT-4 is, however, 1000 times larger in size (170 trillion parameters vs. 175 billion parameters), hinting at enhanced capabilities and accuracy~\cite{koubaa2023gpt}.
In our case, the additional number of tokens available for GPT-4 is welcome since the prompts of the solution are quite large since they include many web elements, encoded in JSON format, with the prompt.
We expect each JSON representation of one web element to be close to 1K characters, meaning that each prompt, with ten web elements, would constitute around 10K characters or 2500 tokens (1 token $\sim$ 4 characters).
This size is also feasible when using GPT-3.5-turbo since it is less than the allotted 4K tokens per prompt.
However, since a JSON structure includes many special characters and digits, we expect a lower ratio than four characters per token (i.e., lower than the expected ratio for pure text).
A ratio of two characters per token results in 5K tokens for the ten JSON representations alone, motivating our selection of GPT-4 that can receive 8K tokens in one prompt.
In conclusion, we choose to use GPT-4 in our experiment even with the drawback of a higher cost, lower RPM (i.e., requests per minute), and lower TPM (i.e., tokens per minute) since increased accuracy and max number of tokens are more important for our evaluation.

\begin{table}[H]
\centering
\caption{Comaprison between OpenAI GPT-versions.}
\label{llms}
\begin{tabular}{|l|l|l|l|l|}
\hline
\textbf{GPT-version} & \textbf{Max tokens} & \textbf{RPM} & \textbf{TPM} & \textbf{Cost 1K tokens} \\ \hline
GPT-3.5-turbo & 4K & 3500 & 90000 & \$0.002 \\ \hline
GPT-4 & 8K & 200 & 40000 & \$0.03 \\ \hline
\end{tabular}
\end{table}

\subsection{Prompt Engineering}

Prompt engineering is the intentional construction and refinement of prompts used in natural language processing tasks.
It involves formulating precise instructions or queries to produce desired responses from large language models.

We experimented with larger and smaller prompts with or without examples to maximize the correctness of the output while trying to keep the prompt length short enough to be of practical use (i.e., since the prompt size is limited and is associated with a cost).

Initially, the experiment was performed with a minimal prompt with no examples (zero-shot).
Hence, each prompt only contained instructions, the ten web element candidates, and a target element in JSON format.
Each JSON element contains the property names and values of one web element.
We created the JSON elements from the following properties: Tag, Visible Text, Class, Id, Name, HRef, Location, Area, Shape, Alt, Is Button, XPath, and Neighbor Text.
Each candidate is also given a unique id to make it possible to ask GPT-4 to return with the id instead of the entire JSON element.
The prompt asks GPT-4 to return the id of the candidate that is most similar to the target web element (also provided in JSON format) and specify a list of reasons for the decision.
Listing~\ref{listing:json} shows an example of such a prompt, including the response from GPT-4.

Next, we reran the experiment with a more descriptive prompt that contained one set of example inputs (one-shot) and the corresponding output.
The one-shot approach was hypothesized to help train the LLM in how to perform the comparison and thereby provide a better result.

Table~\ref{table:results_on_failed} presents our findings from evaluating the zero- and one-shot approaches. 
These were calculated on a subset of 70 web element pairs where VON Similo failed to identify the correct target (i.e., by running VON Similo in all the 804 cases).
These cases were chosen because they were perceived of higher complexity since the conventional algorithm failed to identify them.
As can be seen from the last column in the Table, including one example improved the result from 37 to 41 (i.e., 52.9\% to 58.6\%), representing a 5.7 percent reduction in not located web elements.
Based on this result, and since the additional data for the one-shot did not significantly extend the prompts' token size, we decided to include one example in all the prompts used in the full experiment, i.e., all 804 web element pairs.

\begin{table}[H]
\centering
\caption{The number of located (and not located) web elements when using one or zero examples included in the prompt.}
\label{table:results_on_failed}
\begin{tabular}{|l|l|l|l|l|}
\hline
\textbf{Type} & \textbf{Total} & \textbf{Located} & \textbf{Not located} & \textbf{\% Located} \\ \hline
Zero-shot & 70 & 37 & 33 & 52.9 \\ \hline
One-shot & 70 & 41 & 29 & 58.6 \\ \hline
\end{tabular}
\end{table}

To improve the results even further, we tried to increase the number of candidates sent to the LLM (i.e., a larger list of top candidates proposed by VON Similo).
We observed several drawbacks with increasing the number of candidates: (1) increased cost due to a larger prompt, (2) failure to identify the most similar web element due to many candidates, and (3) GPT-4 needed more detailed examples sticking to the instructed output format (i.e., got confused by the increased prompt size and did not return with the widget id and motivations in the format specified by the prompt).
Instead of exhaustively exploring (i.e., with a different number of candidates), we decided that ten candidates and one set of examples (one-shot) would be sufficient for our experiment.
Thus, concluding that finding an optimal balance of the number of elements is out of scope for this study.
The impact of this design choice results in 13 cases where the correct web element (i.e., according to our oracle) was not part of the top ten candidates sent to the LLM.
Hence, making it impossible for the LLM to select the correct web element.
As a result, by increasing the prompt size, VON Similo LLM could, theoretically, have reported 13 more identified web elements in this study.
However, even doubling the number of candidates from VON Similo (i.e., from ten to 20) would have only resulted in five more instances where the correct element would have been part of the list of widgets sent to the LLM.
As such, we concluded that the additional results would not outweigh the additional prompt size and cost of using the GPT-4 API.

\begin{lstlisting}[style=DOS, breaklines=true, numbers=none, caption={Sample GPT-4 prompt with response.}, label=listing:json]
Given the following candidate web elements (|| means that an attribute can have multiple values):
{widget_id:"202",tag:"a",text:"Beauty, Health",href:"https://www.aliexpress.com/category/66/health-beauty.html",location:"277,541",area:"1584",shape:"488",is_button:"no",xpath:"/html/body/div/div[5]/div/div[2]/div/div[2]/div/div[2]/dl[11]/dt/span/a",neighbor_text:"toys kids & babies outdoor fun sports beauty health hair automobiles motorcycles home improvement tools"}
{widget_id:"200",tag:"span || a",text:"Outdoor Fun & Sports",href:"https://www.aliexpress.com/category/18/sports-entertainment.html",location:"236,497",area:"8400",shape:"685",is_button:"no",xpath:"/html/body/div/div[5]/div/div[2]/div/div[2]/div/div[2]/dl[10]/dt/span",neighbor_text:"bags & shoes toys kids babies outdoor fun sports beauty health hair automobiles motorcycles"}
{widget_id:"201",tag:"span",text:"Beauty, Health & Hair",location:"236,532",area:"8400",shape:"685",is_button:"no",xpath:"/html/body/div/div[5]/div/div[2]/div/div[2]/div/div[2]/dl[11]/dt/span",neighbor_text:"toys kids & babies outdoor fun sports beauty health hair automobiles motorcycles home improvement tools"}
{widget_id:"204",tag:"span || a",text:"Automobiles & Motorcycles",href:"https://www.aliexpress.com/category/34/automobiles-motorcycles.html",location:"236,567",area:"8400",shape:"685",is_button:"no",xpath:"/html/body/div/div[5]/div/div[2]/div/div[2]/div/div[2]/dl[12]/dt/span",neighbor_text:"outdoor fun & sports beauty health hair automobiles motorcycles home improvement tools"}
{widget_id:"197",tag:"span",text:"Toys , Kids & Babies",location:"236,462",area:"8400",shape:"685",is_button:"no",xpath:"/html/body/div/div[5]/div/div[2]/div/div[2]/div/div[2]/dl[9]/dt/span",neighbor_text:"home pet & appliances home bags shoes toys kids babies outdoor fun sports beauty health hair"}
{widget_id:"199",tag:"a",text:"Kids & Babies",href:"https://www.aliexpress.com/category/1501/mother-kids.html",location:"313,471",area:"1476",shape:"455",is_button:"no",xpath:"/html/body/div/div[5]/div/div[2]/div/div[2]/div/div[2]/dl[9]/dt/span/a[2]",neighbor_text:"home pet & appliances bags shoes toys kids babies outdoor fun sports beauty  health hair"}
{widget_id:"194",tag:"span",text:"Bags & Shoes",location:"236,427",area:"8400",shape:"685",is_button:"no",xpath:"/html/body/div/div[5]/div/div[2]/div/div[2]/div/div[2]/dl[8]/dt/span",neighbor_text:"jewelry & watches home pet appliances home bags shoes toys kids babies outdoor fun sports"}
{widget_id:"206",tag:"a",text:"Home Improvement",href:"https://www.aliexpress.com/category/13/home-improvement.html",location:"277,611",area:"2196",shape:"677",is_button:"no",xpath:"/html/body/div/div[5]/div/div[2]/div/div[2]/div/div[2]/dl[13]/dt/span/a",neighbor_text:"beauty health & hair automobiles motorcycles home improvement tools"}
{widget_id:"205",tag:"span",text:"Home Improvement & Tools",location:"236,602",area:"11040",shape:"521",is_button:"no",xpath:"/html/body/div/div[5]/div/div[2]/div/div[2]/div/div[2]/dl[13]/dt/span",neighbor_text:"beauty health & hair automobiles motorcycles home improvement tools flash deals"}
{widget_id:"203",tag:"a",text:"Hair",href:"https://www.aliexpress.com/category/200002489/hair-extensions-wigs.html",location:"381,541",area:"450",shape:"138",is_button:"no",xpath:"/html/body/div/div[5]/div/div[2]/div/div[2]/div/div[2]/dl[11]/dt/span/a[2]",neighbor_text:"toys kids & babies outdoor fun sports beauty  health hair automobiles motorcycles home improvement tools"}

find the one that is most similar (answer with the widget_id of the most similar and motivate why using a list) to the element:
{tag:"span || a",text:"Health & Beauty",href:"https://www.aliexpress.com/category/66/health-beauty.html",location:"237,465",area:"7140",shape:"793",is_button:"no",xpath:"/html/body/div/div[4]/div/div/div[2]/div/div[2]/dl[11]/dt/span",neighbor_text:"toys, kids & baby toys sports outdoors health beauty automobiles motorcycles home improvement"}

The most similar element is the one with widget_id "201". The reasons for this choice are:

1. Both elements have "span" as one of their 'tag' attribute.
2. The text "Beauty, Health & Hair" in the element with widget_id "201" is closely related to the text "Health & Beauty" in the given element.
3. The 'location', 'shape', 'is_button', and 'neighbor_text' attributes in both elements have similar values, indicating that they might be close to each other on the layout of the website and have a similar structure.
4. Although the given element has an 'href' attribute and the element with widget_id "201" does not, this could be a minor change during the evolution of the web application, and the overall similarity of other attributes makes it the best candidate.
\end{lstlisting}

\subsection{Locating Web Elements} \label{Locating Web Elements}

We created a tool (implemented in Java and included in the replication package) that uses the extracted web element properties (see Section \ref{Extracting}) to compare the effectiveness and efficiency of the two approaches.

\begin{figure}[H]
  \centering
  \includegraphics[width=0.8\textwidth]{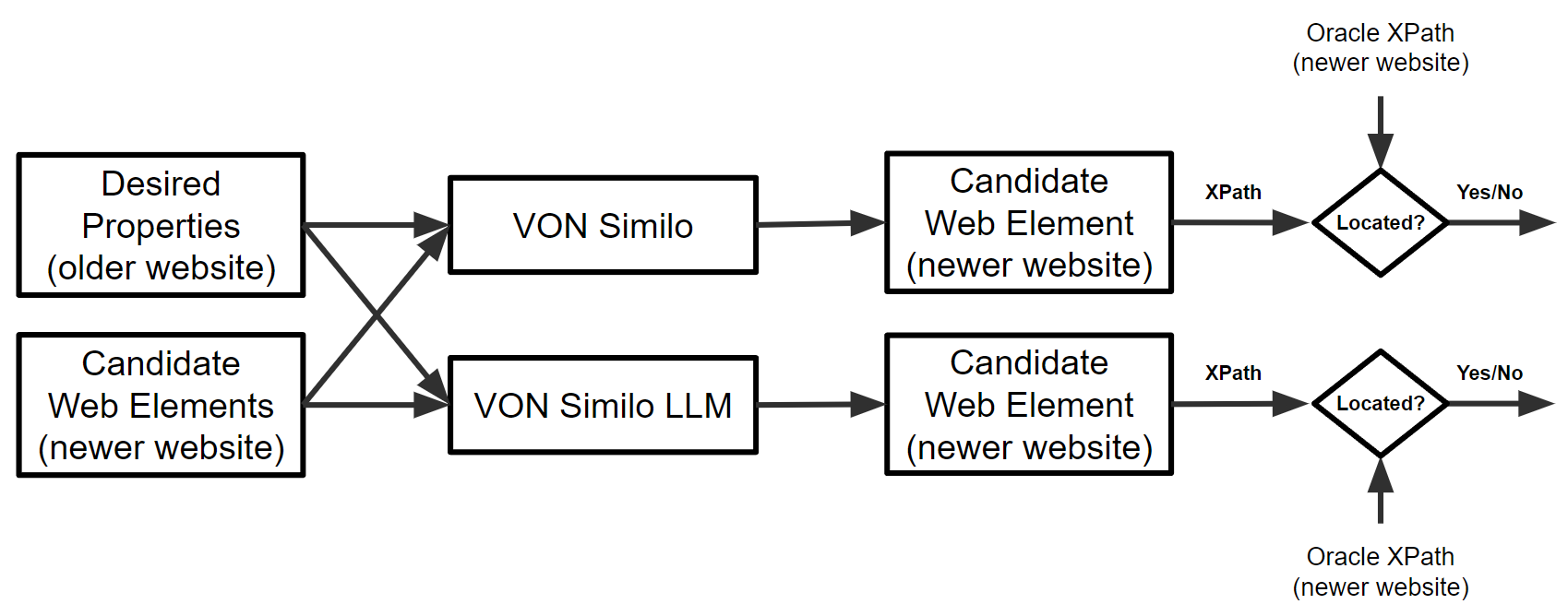}
  \caption{The process of locating a candidate web element from desired properties using the two approaches.}
  \label{locatingwebelements}
\end{figure}

Figure~\ref{locatingwebelements} shows the process of locating a candidate web element among the candidates extracted from the newer web application version based on the target's desired properties extracted from an older version of the same application.
For each of the 804 web elements that were previously extracted from the older versions of 48 web applications, the desired properties and all the available candidates for the web application homepage were submitted as input to both approaches.
VON Similo and VON Similo LLM then identify the candidate that holds properties most similar to the desired properties by comparing the properties of each candidate.
Next, the XPaths of an identified candidate are compared with the Oracle XPath.
Note that each candidate can have multiple XPaths due to the VON concept since a visual web element may consist of several DOM elements.
The candidate is considered located if any of the candidate XPaths are identical to the Oracle XPath (and not located otherwise).
Table~\ref{table:localizationresult} contains a summary of the two possible outcomes after a localization attempt.

\begin{table}[H]
\centering
\caption{Description of the localization result.}
\label{table:localizationresult}
\begin{tabular}{|l|p{10cm}|} 
\hline
\textbf{Localization result} & \textbf{Description} \\ \hline
Located & The approach is able to identify the correct candidate web element where one of the XPaths is identical to the oracle. \\ \hline
Not located & The approach finds a match among the candidate web elements, but none of the XPaths are identical to the oracle. \\ \hline
\end{tabular}
\end{table}

\begin{figure}[H]
  \centering
  \includegraphics[width=0.6\textwidth]{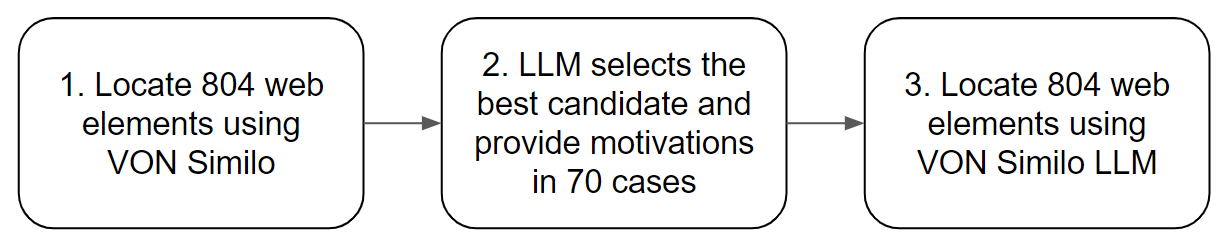}
  \caption{Overview of the three phases of the experiment.}
  \label{phases}
\end{figure}

We decided to divide the experiment into three phases.
Figure \ref{phases} contains an overview of the phases further explained below.
The first and last phases target research questions RQ1 and RQ2, while we aim to answer RQ3 with results from the second phase.
\begin{enumerate}
\item 
Initially, we attempted to locate all the 804 web elements using VON Similo, which resulted in 70 not being found (see Section \ref{Results}) in the newer web application versions based on the properties extracted from older versions.

\item 
Next, we asked the LLM (i.e., GPT-4) to identify the correct web element and motivate that choice, given the ten top-ranked elements provided by VON Similo, for the 70 cases where VON Similo failed.
We analyzed the motivations given by the LLM to tell if the motivations were based on semantic understanding, context awareness, or using a standard comparison operator (i.e., like VON Similo). See definitions in Section \ref{Results}.
The three categories were coded based on literature that utilizes abilities in traditional algorithms, NLP, or LLMs when comparing GUI elements or creating input for testing \cite{nass2022similarity, nass2023robust, liu2022fill, liu2023chatting, wu2023screen}.

We decided to use a subset of the 804 cases to lower the cost of using the LLM API and to reduce the number of motivations to categorize.
Selecting the cases where VON Similo failed has several benefits: (1) it is a significantly smaller sample (i.e., less costly), (2) the correct alternative is never the first candidate (i.e., since VON Similo failed), making the choice less evident, and (3) it is more valuable if the LLM can find the correct web element when the conventional approach (i.e., VON Similo) fails.

\item 
Finally, to evaluate VON Similo LLM, we extracted the top ten elements that best match all of the 804 oracles (i.e., correct targets) using VON Similo and asked the LLM to select the candidate that is most similar to the oracle (i.e., the properties extracted from the older version) for all oracles.
To optimize (i.e., reduce) the cost and time of the experiment, we did not ask the LLM to provide us with motivations, instead only to return with the id of the best candidate. 
This design greatly reduced the output prompt from the LLM and, thereby, the execution time since each output character increases the execution time and cost of using the LLM API.
\end{enumerate}


\section{Results}\label{Results}

In this section, we present the results of the experiment study. 
We present the results according to the order of the study's three research questions.

\subsection{RQ1 - Effectiveness}

Table~\ref{table:results_approaches} contains the result when comparing the effectiveness, in terms of being able to locate the correct candidate based on desired properties, of the two approaches.
When attempting to identify the correct candidate in 804 cases extracted from 48 web applications, VON Similo failed to locate the correct candidate in 70 cases (i.e., 91.3\% correctly located).
In comparison, the VON Similo LLM approach (i.e., use an LLM to identify the best candidate among the ten provided by VON Similo) only failed in 40 cases (i.e., 95.0\% correctly located).
Thus, resulting in a 42.9 percent reduction of not-located web elements when using VON Similo LLM.

\begin{table}[H]
\centering
\caption{The total number of located (and not located) web elements for the two approaches.}
\label{table:results_approaches}
\begin{tabular}{|l|l|l|l|l|l|l|}
\hline
\textbf{Approach} & \textbf{Total} & \textbf{Located} & \textbf{Not located} & \textbf{\% Located} & \textbf{API Cost (\$)} & \textbf{Time/localization (ms)} \\ \hline
VON Similo & 804 & 734 & 70 & 91.3 & 0 & 29 \\ \hline
VON Similo LLM (one-shot)& 804 & 764 & 40 & 95.0 & 35.86 & 1934 (STD 537) \\ \hline
\end{tabular}
\end{table}

\begin{figure}[H]
  \centering
  \includegraphics[width=0.5\textwidth]{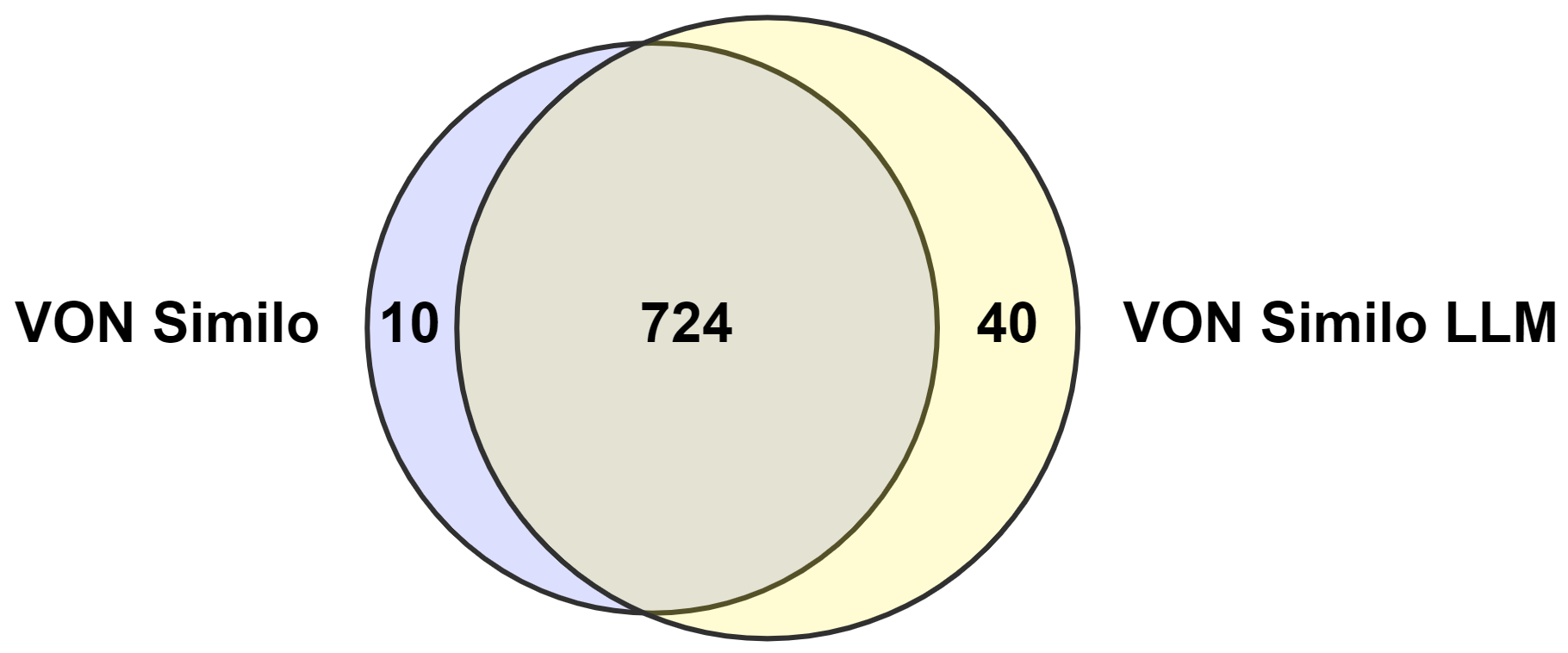}
  \caption{Venn diagram containing the number of correctly located candidates (i.e., web elements) for each approach.}
  \label{venn}
\end{figure}

The Venn diagram in Figure~\ref{venn} shows the number of located web elements by VON Similo and VON Similo LLM.
Both approaches located 724 of the correct candidates.
The VON Similo LLM approach located 40 candidates that VON Similo did not locate, and VON Similo located ten candidates that VON Similo LLM failed to locate.

Because we instructed the LLM to only provide a widget id as output and no motivation, in this experiment, it is impossible to analyze why VON Similo LLM did not find the ten cases VON Similo found (further elaborated on in Section \ref{Discussion}).

\begin{center}\fbox{
\begin{minipage}[t]{0.96\linewidth}
To summarize, for what concerns research question RQ1, using the VON Similo LLM approach instead of the conventional VON Similo algorithm, we reduced the number of not located candidates from 70 to 40 cases, i.e., 42.9 percent. 
\end{minipage}
}
\end{center}

\subsection{RQ2 - Efficiency}

The Time/localization column in Table~\ref{table:results_approaches} shows the average time in milliseconds to locate one candidate using both approaches (29 vs. 1934 ms).
Also, within parentheses, the standard deviation is included for the VON Similo LLM approach (537 ms).
We were unable to measure the standard deviation of the VON Similo approach due to the lack of precision (i.e., we could only measure whole milliseconds).
As expected, the performance of the VON Similo algorithm is much higher (i.e., almost two magnitudes lower execution time) than the VON Similo LLM approach due to the slow response time of the GPT-4 API.

\begin{center}\fbox{
\begin{minipage}[t]{0.96\linewidth}
To summarize research question RQ2, the performance of the LLM approach at the time of writing is almost two magnitudes slower than the conventional algorithm due to the long response time from the GPT-4 API (i.e., around 2 seconds on average).
While we cannot generalize this result to all LLM solutions, it gives insights into a snapshot of the order of magnitude of time required when we conducted this study.
\end{minipage}
}
\end{center}

\subsection{RQ3 - What main aspects does a large language model use when comparing similar web elements when a conventional approach struggles?}

\begin{figure}[H]
  \centering
  \includegraphics[width=0.6\textwidth]{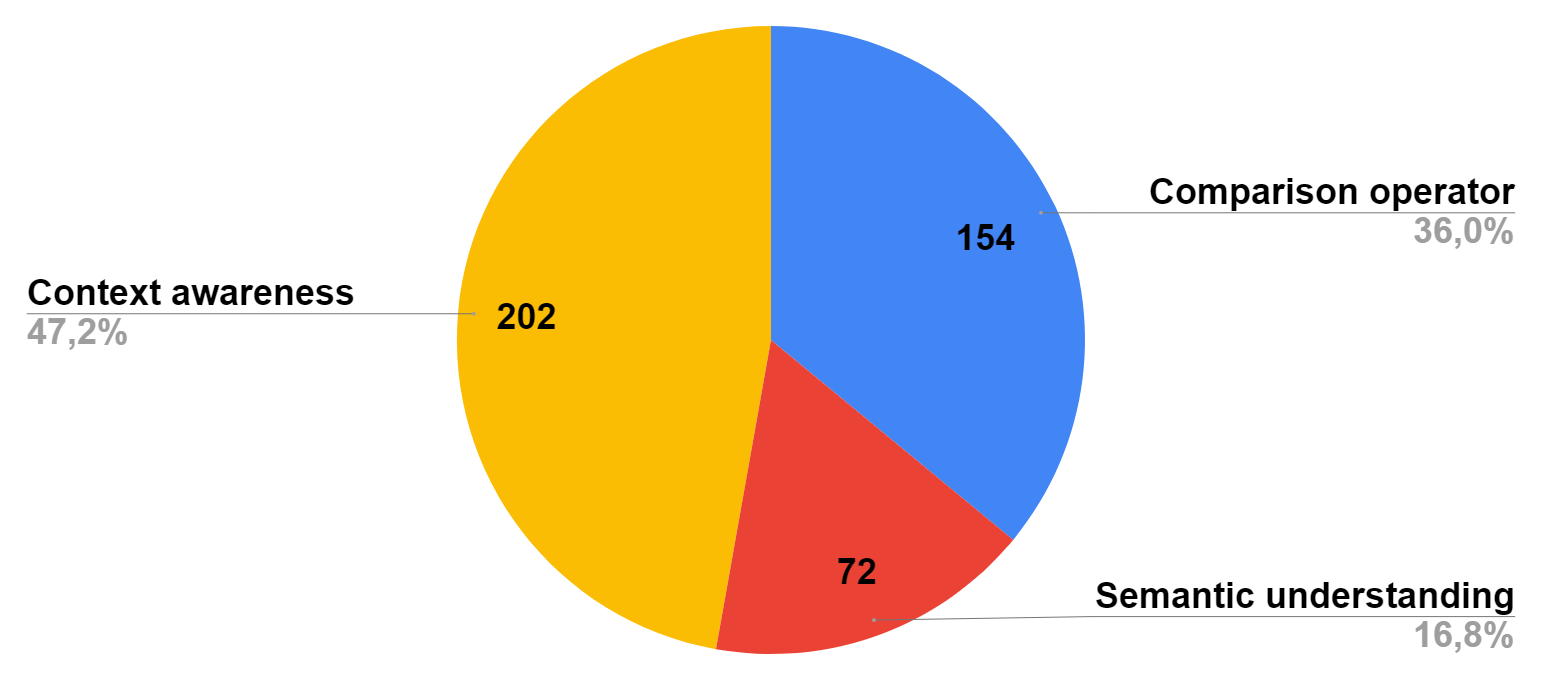}
  \caption{Motivations from the LLM classified as codes.}
  \label{pie}
\end{figure}

Figure~\ref{pie} shows the results from our qualitative content analysis of the 428 motivations provided by GPT-4 for all the 70 cases (i.e., six motivations per case, on average) when VON Similo could not identify the correct candidate.
We provided motivations for selecting the cases where VON Similo failed in Section \ref{Locating Web Elements}.

We defined three categories of motivations before the analysis (see Section \ref{Locating Web Elements}) to be able to evaluate how frequently GPT-4 incorporates either of the aspects; comparison operator, semantic understanding, or context awareness, in its motivations:

\begin{itemize}
\item \textbf{Comparison operator:} Motivation based on conventional comparison operators (e.g., equals, Euclidean distance, Levenshtein distance).
Hence, the only category that the VON Similo approach uses to identify elements.
\item \textbf{Semantic understanding:} Motivation based on semantic understanding. Semantic understanding refers to interpreting the meaning of information within its context. It involves understanding the relationships between words, sentences, and concepts and the intended or implied meaning behind them.
\item \textbf{Context awareness:} Motivation based on context awareness. Context awareness refers to the capability to perceive and understand the situational context (e.g. layout and positioning of elements in web applications, in our case).
\end{itemize}

We categorized 202 motivations (i.e., 47\%) to be associated with context awareness, 72 motivations (i.e., 17\%) to be associated with semantic understanding, and 154 motivations (i.e., 36\%) to be associated with the use of some form of conventional comparison operation (e.g., equals).
Some motivations could belong to more than one category.
In those cases, we sorted the motivation into the nearest category (i.e., the most appropriate according to the authors).
We did not encounter any motivations leading us to refine existing, or add new, categories while performing the analysis.
Table~\ref{table:example_motivations} contains examples of motivations returned from GPT-4, categorized as associated with the comparison operator, semantic understanding, or context awareness.
Half of the motivations were gathered from the first prompt responses, while the remaining examples were manually selected to show some alternate or interesting motivations.
When comparing motivations from GPT-4 when it was correct or incorrect (i.e., selected the correct candidate according to the oracle), we did not find any pattern in the motivations that would indicate when it was more or less confident of the selection of the most similar candidate.

\begin{center}\fbox{
\begin{minipage}[t]{0.96\linewidth}
To summarize research question RQ3, the result indicates that the LLM mainly uses context awareness or semantic understanding (64\% of the time) rather than relying on some form of comparison operation (i.e., like a conventional, non-AI algorithm) in more complex cases where a conventional approach failed.
\end{minipage}
}
\end{center}


\begin{table}[H]
\centering
\scriptsize{
\caption{Example motivations from GPT-4 classified as comparison operator, semantic understanding, or context awareness.}
\label{table:example_motivations}
\begin{tabular}{|p{54mm}|p{54mm}|p{54mm}|}
\hline
\textbf{Examples that indicate that Comparison operator is used to find the target} & \textbf{Examples that indicate that Semantic understanding is used to find the target} & \textbf{Examples that indicate that Context awareness is used to find the target} \\ \hline
Both elements have "span" as one of their 'tag' attribute. & The text "Beauty, Health \& Hair" in the element with widget\_id "201" is closely related to the text "Health \& Beauty" in the given element. & The 'location', 'shape', 'is\_button', and 'neighbor\_text' attributes in both elements have similar values, indicating that they might be close to each other on the layout of the website and have a similar structure. \\ \hline
Both elements have "a" or "span" as their 'tag' attribute. & The text "Sign up" in the element with widget\_id "8065" is closely related to the text "Log in" in the given element, as both texts are related to account actions. & Although the given element has an 'href' attribute and the element with widget\_id "201" does not, this could be a minor change during the evolution of the web application, and the overall similarity of other attributes makes it the best candidate. \\ \hline
The 'class' attribute values in both elements are very similar, containing "nav-logo-base" and "nav-sprite". & The text "Order Status" in the element with widget\_id "1823" is not exactly the same as the text "Shopping History" in the given element, but both texts are related to account and order information, which leads to the assumption that they are similar in purpose. & Both elements have a similar 'location' attribute, indicating that they might be close to each other on the layout of the website. \\ \hline
The 'href' attribute in both elements is the same, as they both point to the same URL ("https://www.cnn.com/us"). & The text "Account" in the element with widget\_id "1815" is not exactly the same as "Store Locator" in the given element, but there's no other candidates with the text "Store Locator". In this case, "Account" may represent a location-related functionality. & Both elements have relatively large 'area' and 'shape' attributes, suggesting that they are both prominent elements on the webpage. \\ \hline
The 'location' attribute in both elements is the same: "20,20". & The text "Upgrade to premium" in the element with widget\_id "8817" is closely related to the text "Get premium" in the given element. & The 'location' attribute indicates that they might be far apart in the layout of the website, but the 'neighbor\_text' attribute has some overlapping words (e.g., "spotify", "support", "download", "premium"). \\ \hline
The text "Enterprise" is exactly the same in both elements. & The text "Start your free trial" in the element with widget\_id "3214" is closely related to the text "Try free for 30 days" in the given element. & Both elements have a similar 'location' attribute with only a minor difference in the x coordinate, indicating that they are situated near each other on the layout of the website. \\ \hline
The 'id' attribute in both elements is the same: "hero-banner-get-office-link". & The text "Support" in the element with widget\_id "10880" is closely related to the text "Help" in the given element. Both serve the same purpose of assisting users with issues or questions. & Despite some differences in 'xpath', both elements seem to be part of the navigation menu, as suggested by the 'neighbor\_text' attribute. \\ \hline
The text "Find jobs" in the element with widget\_id "7973" is identical to the text "Find Jobs" in the given element. & The 'neighbor\_text' attribute is similar in both elements, with both mentioning social platforms like "twitter", "instagram", "snapchat", "youtube", and "the espn daily podcast". & The text "Items in cart" in the given element is related to the functionality of a shopping cart, and the element with widget\_id "12341" also has a cart-related functionality, although the text is not present. \\ \hline
Both elements share the same 'href' attribute, which points to "https://www.instructure.com/". & The text "Claims Support" in the element with widget\_id "11882" is closely related to the text "Delivery Issues" in the given element, as both deal with issues regarding deliveries. & The 'xpath' and 'neighbor\_text' attributes also show similarities, suggesting that they are part of the same group of links within the footer of the website. \\ \hline
The text "Cart" is present in both elements. & The text "Plans \& Pricing" in the element with widget\_id "13858" is closely related to the text "PLANS" in the given element. & Although the 'href' attribute is different, the change could be due to the updated web application using a different method to handle account sign-in functionality. \\ \hline
\end{tabular}
}
\end{table}

\section{Discussion}\label{Discussion}

LLMs with human-like abilities such as semantic understanding and context awareness have the potential to increase the effectiveness of identifying web elements.
Instead of just comparing attributes and other properties (i.e., like a conventional algorithm), LLMs can relate to the meaning of neighbor texts, understand the purpose of an element, and evaluate the structure (i.e., both the DOM and visually in terms of layout and element placement) to make more informed decisions when comparing and identifying web elements. 
One example is the following motivation from the LLM: "The 'location' attribute indicates that they might be far apart in the layout of the website, but the 'neighbor\_text' attribute has some overlapping words (e.g., 'spotify', 'support', 'download', 'premium'.)".
This and the following examples can be found in Table~\ref{table:example_motivations}.
LLMs recognize common patterns such as menus, forms, footers, or groups and use this contextual information to refine the identification process.
For example, the LLM motivated one decision with the text: "The 'xpath' and 'neighbor\_text' attributes also show similarities, suggesting that they are part of the same group of links within the footer of the website.".
Another example is: "Despite some differences in 'xpath', both elements seem to be part of the navigation menu, as suggested by the 'neighbor\_text' attribute.".
With almost human-like abilities when identifying web elements, LLMs can reduce the need for manual intervention and script maintenance in tools and frameworks for web-based test automation.
More reliable test scripts save time for the human testers, who can focus on more meaningful tasks like test strategies and exploratory testing.

There is also a downside to utilizing GPT-4 (i.e., the LLM used in our experiment) for web element identification.
API requests are very slow today compared to a conventional algorithm like VON Similo.
We measured the average API request to be around two seconds, which would result in a noticeable delay even in an automated GUI script (i.e., that, in turn, is very slow compared to Unit tests).
Although we expect future advancements of GPT and other LLMs to become faster, there might always be some delay that would affect the execution time of the automated test script in a noticeable way.

Using GPT-4 also comes with a cost in terms of a fee charged by OpenAI for utilizing the API.
The cost is not easy to grasp since it is based on the number of tokens sent between the client and server.
According to our measurements, see Table~\ref{table:results_approaches}, the cost is not negligible (\$36 for 804 prompts, i.e., \$0.045 per prompt) and needs to be taken into consideration when evaluating if the price of using the API (i.e., runtime cost) is lower than the expected reduction in maintenance cost.
Such a calculation is complicated due to the many variables that affect the maintenance cost (e.g., software maturity, time between releases, number of test cases, size of the test cases, and the salary of developers).
However, assuming a test suite with an average maintenance time of 110 minutes per test case between two major versions, 47 localizations on average per test case, and an estimated cost of 100 dollars per hour for an employee (as reported in Alégroth et al.~\cite{alegroth2016maintenance}), the LLM approach would likely provide a positive return on investment in just one software release cycle.
The cost of test case maintenance would be: 100 * (110 / 60) = \$183, and the cost of using the GPT-4 API would be: 47 * 0.045 = \$2.
Since the cost of utilizing the GPT-4 API is negligible compared to the manual maintenance cost, the additional robustness gained by the LLM approach would likely be more valuable.
Assuming the same increase in robustness as in our results (40 vs. 70 not located), the maintenance cost would be reduced to: 183 * (40 / 70) = \$105.
Even though the calculations are based on industrial data, they only provide an indication since salaries for engineers differ globally.

However, the cost and payment plans will likely also change over time.
There may even be adequate LLMs that are entirely free or that you can install locally, eliminating, at least, the cost aspect.
A locally installed LLM would probably also impact the performance but may not eliminate the problem that the API request delay has a noticeable effect on the test execution.
That GPT-3.5-turbo is considerably faster than GPT-3 is also an indication that we might expect to see a turbo version of GPT-4 in the future.

Since LLMs are based on artificial neural networks (ANNs) \cite{jain1996artificial}, we can only assume that the motivations provided by the LLM have anything to do with the candidate selected since a large ANN can be seen as a black box model (i.e., with inputs and outputs) that we cannot fully comprehend due to the complexity of the network.
Ongoing discussion exists about whether LLMs are probabilistic models or if they truly learn to understand the world~\cite{y2022large, min2022rethinking, razeghi2022impact}.

In summary, LLMs can be used to further improve web element localization due to their assumed semantic understanding and context awareness with the drawbacks of slower test execution and the cost of using the API.
However, more research is needed to fully grasp the potential and shortcomings of using LLMs for web element identification and if the models actually possess knowledge about context and semantics.

\section{Threats to Validity}
To reduce the internal validity threat, we limited the influence of the selection of web elements on our experiment by selecting specific categories of web elements that could be used for actions, assertions, or synchronization and that were available on both versions of the website's homepage.
As we focused on investigating the web element finding ability only, we do not believe that the possibility that elements on the homepage differ significantly from other web elements is a substantial threat.

The choice of applications and versions analyzed in the study may compromise its external validity.
To address this issue, we opted to focus on the top 48 sites based on Alexa.com rankings, as we have no control over the websites listed on that site.
Additionally, the version of a website can impact the number of failed localization attempts, mainly since longer intervals between releases often result in more changes.
To mitigate this concern, we selected the same interval (one to five years) for website versions as previous studies conducted by Leotta et al.~\cite{leotta2015using} and Nass et al.~\cite{nass2022similarity}.

The construct validity is low since the time between releases (i.e., between 12 to 60 months) should be compared with a typical case in the industry.
However, industrial cases differ a lot.
Some test suites are run every time the source code is updated (i.e., several times per day), while some test suites are run with months in between.
We decided to prioritize getting some changes (i.e., both larger and smaller) by picking a greater period between releases to reduce the risk of not finding any changes at all.

That we selected to use GPT-4 from OpenAI in favor of some other LLM can impact the construct validity since choosing a different LLM would likely give a different result.
We tried to mitigate this threat by motivating our selection of LLM when focusing on effectiveness before efficiency and cost.
Also, future LLM versions will likely be even more capable, making this study merely a baseline for the future.

Limiting the number of candidates to ten was made to prevent the prompt from exceeding the quota and reduce the API's runtime cost.
However, this design choice leads to 13 cases where none of the top ten candidates are correct, resulting in inevitable failure for the LLM.
This threat to validity arises because the chosen constraint on the number of candidates potentially restricts the LLM's ability to provide accurate responses.
By limiting the available options, the experiment does not fully assess the LLM's ability to generate appropriate and correct responses.
This limitation could lead to an underestimation of the LLM's performance, as it may have the potential to generate correct responses beyond the limited set of candidates.
An alternative approach to mitigating this threat would have been to increase the number of candidates to 20, which would have included additional correct candidates in the top 10.
However, concerns about the impact on prompt size and cost led to the decision against this option.

\section{Related Work}\label{Related Work}

In practical terms, two categories of methodologies have emerged, each possessing contrasting yet non-contradictory characteristics: post-repair approaches that address locator failures by employing remedial measures and more preventive strategies that focus on generating resilient locators.
Only a few of the current algorithms and approaches utilize natural-language processing (NLM) or large language models (LLMs).
This Section covers them both, emphasizing the ones taking advantage of LLM or NLM.

\subsection{Post-repair approaches}

This category of approaches aims to automatically repair the automated test execution or script after a failure has occurred (i.e., post-execution).
Automatic repair reduces the costly manual labor of repairing test cases or scripts and has been researched by many, e.g.,~\cite{choudhary2011water, 10.1145/2950290.2950294, Kirinuki2019COLORCL}.

Khaliq et al. \cite{khaliq2022transformers} proposed a novel automated GUI testing approach using a sequence-to-sequence transformer model in GPT-2, which perceives the application state through element classification and generates test flows in English.
Their model aims to repair flaky tests when the GUI is modified and automatically generate new test flows for regression without manual intervention.
They showed that abstract English test flows could be converted into executable test scripts using a simple parser.

A more conventional approach (i.e., non-AI), named WATER, proposed by Choudhary et al.~\cite{choudhary2011water}, compares the test execution on two software versions, one where the test succeeds and one when it fails.
In common with Similo (and VON Similo), WATER uses weighted locator parameters when repairing a broken locator.
The WATER approach is, however, a post-repair technique and utilizes an entirely different set of locator parameters than Similo (i.e., XPath, coord, clickable, visible, index, and hash) that are compared using equality or Levenshtein distance~\cite{levenshtein}.

Another post-repair tool is WATERFALL~\cite{10.1145/2950290.2950294}.
WATERFALL is an advancement on WATER and uses the same heuristics for repairs but can improve the effectiveness of script repair (by 209\%) by taking advantage of the knowledge that minor versions occur between major versions in software releases.

COLOR, proposed by Kirinuki et al.~\cite{Kirinuki2019COLORCL}, is another approach that uses several attributes, positions, images, and other properties to suggest a repair.
Their experiments show that COLOR can be more effective than WATER (especially concerning more complex changes, like switching from one web page to another) and that the algorithm can identify the repair with 77\% to 93\% accuracy.

Repairing broken locators utilizing a DOM tree comparing algorithm is an approach presented by Brisset et al.~\cite{brisset2022erratum}.
They compared their tool, Erratum, with WATER and found that it has 67\% higher accuracy.

Grechanik et al.~\cite{grechanik2018differencing,xie2009guide} proposed GUIDE, a tool for a non-intrusive, platform-, and language-independent repair algorithm for web applications by identifying changes occurring between two released software versions.
The tool can be used for suggesting repairs or providing guidance for test planning.

\subsection{Resilient locators}

Resilient (i.e., robust) locators in GUI test automation refer to the challenge of reliably identifying and interacting with GUI elements during automated testing.
Changes in GUI layout and dynamic content can cause locators to fail, leading to test script failures.
Researchers aim to develop techniques for generating robust locators tolerant to GUI changes, ensuring efficient and reliable test automation.
Many approaches have been proposed seeking to mitigate this problem in the literature.

A study by Kirinuki et al. attempts to solve the locator maintenance problem by not relying on attributes and the structure of the DOM and instead leverages NLP with heuristic search to identify web elements in web pages from natural-language-like test cases \cite{kirinuki2022web}.
An example of such a test step could be: enter 'admin' in 'username'.
Evaluation of three open-source web applications showed a success rate of 94\% in identifying web elements and correct identification in 68\% of the test cases.

Another interesting approach that takes advantage of GPT (i.e., GPT-3 in this case) while avoiding the shortcomings of a traditional test script is GPTDroid, proposed by Zhe Liu et al. \cite{liu2023chatting}.
Utilizing the strengths of ChatGPT (i.e., understanding human knowledge), they formulate test steps in plain English and pass the GUI page content to the LLM.
Next, the LLM responds with an instruction about what step to do next when asked: 'What operation is required?'.

Zhe Liu et al. also proposed to use the power of an LLM to automatically generate more realistic test scenarios that can interact with a GUI application more similar to a human tester, for example, fill out forms with suitable content that makes it possible to progress to the next step.
Their tool QTypist, can generate text input related to the GUI context and semantic requirement, thereby enabling better test coverage \cite{liu2022fill}.

CrawLabel is a test-generation tool (a plugin for Crawljax) that utilizes grammar learning (i.e., NLP) to perform unsupervised end-to-end testing of web applications \cite{liu2022crawlabel}.

Among the more traditional algorithms (i.e., not utilizing some form of AI), we need to mention the approaches (i.e., Similo and VON Similo) we aim to advance in this paper.
The different Similo approaches are covered, in detail, in Section \ref{Similo algorithm}.

Several approaches attempt to create robust XPath locators.
The algorithm proposed by Montoto et al.~\cite{montoto2011automated} is one of them and uses a bottom-up strategy to generate a change-resilient XPath locator iteratively.
Starting from a simple XPath expression, the algorithm concatenates sub-expressions until the resulting XPath can uniquely identify the target element.
If the resulting XPath is not unique, the attribute values of the ancestors are considered until the root is reached.

Other approaches that generate robust XPaths are ROBULA~\cite{leotta2014reducing} and ROBULA+~\cite{leotta2016robula+}, proposed by Leotta et al.
ROBULA+ improves upon the earlier ROBULA algorithm and is often considered state-of-the-art in generating resilient XPath locators for web applications.
The idea behind ROBULA+ is to generate a short but robust locator as possible, given the content of the web page and heuristics about the robustness of various attributes.
ROBULA and ROBULA+ begin with a generic XPath that selects all the nodes in the DOM (i.e., similar to the Montoto approach).
Next, the algorithms refine the XPath, using a set of transformations or prioritizations until only one element is selected.

While some solutions aim to increase the resilience of XPath locators (e.g., ROBULA+ and Montoto), other approaches increase the number of information sources (e.g., attributes and other properties), thereby introducing voting mechanisms or triangulation when identifying the target web element.
The multi-locator, proposed by Leotta et al., is an example that takes advantage of several locators (i.e., with diverse strengths and weaknesses) and uses a voting procedure to select the best candidate web element (i.e., the top-voted one) \cite{leotta2015using}.

Another interesting approach, ATA-QV, proposed by Yandrapally et al.~\cite{yandrapally2014robust}, is to take advantage of neighboring web elements instead of only relying on attributes and properties of each web element.
We can use the information extracted from neighbor web elements to triangulate the location of the target web element.
For example, assume we have a text field with a label describing the text field on the left and a button on the right side.
Even if the attributes and properties of the text field change entirely from one version to the other, it might still be possible to find it by utilizing the label on the left side and the button's caption on the right side.
ATA-QV is an improvement to the technique and tool called ATA proposed by Thummalapenta et al.~\cite{thummalapenta2012automating}.
ATA is a commercial tool that was developed in collaboration with IBM that aims to increase the resilience of locators by relying more on labels (i.e., visual attributes) than the DOM structure.

Nguyen et al. recently suggested an approach that can generate resilient locators by using a new way of constructing XPaths that relies on semantic structures and neighbor web elements and a rule-based method for selecting the best (i.e., most robust) one \cite{nguyen2021generating}.

SIDEREAL is a tool for automated end-to-end (E2E) testing of web applications~\cite{leotta2021sidereal}.
It addresses the problem of broken locators by using a statistical adaptive algorithm that learns the potential fragility of web element properties to generate robust XPath locators.
Compared to the baselines (i.e., ROBULA+ and Montoto), SIDEREAL significantly reduces the number of broken locators, resulting in more reliable E2E testing for web applications.

There are also some commercial products that can learn and adapt their web element localization from existing applications or application versions, like Testitm \footnote{https://www.testim.io/blog/why-testim/} and Ranorex \footnote{https://www.ranorex.com/blog/machine-trained-algorithm/}.

The Similo approach combines many of the techniques of these related works.
For example, Similo utilizes multiple sources of information like the multi-locator approach by Leotta et al. and triangulating using neighbor web elements like the ATA-QV approach by Yandrapally et al.~\cite{yandrapally2014robust}.

VON Similo LLM enhances standard Similo by adding a semantic understanding of attributes (e.g., the caption) in web elements like the approaches proposed by Kirinuki et al. and Zhe Liu et al. \cite{liu2023chatting}.
However, VON Similo LLM goes beyond the semantic understanding of web elements since GPT-4 displays some form of context awareness by relating to the possible use of web elements in a web page or application, taking it even further than the ATA-QV approach by Yandrapally et al.~\cite{yandrapally2014robust}.

\section{Conclusions}\label{Conclusions}

Accurate web element localization is crucial for robust automated scripts in web-based test automation.
Traditional approaches lack semantic understanding and context awareness.
The emergence of Large Language Models (LLMs) like GPT-4 offers human-like abilities that can enhance web element identification.
This study highlights the potential benefits (but also challenges) of using LLMs for web element localization in an automated GUI test case.
Our results show that LLMs can be employed to understand the purpose of elements, analyze neighboring text, and evaluate web page structures, enabling more accurate localizations.
They can reduce manual intervention and script maintenance, freeing human testers' time for more meaningful tasks.
However, using LLMs through APIs like GPT-4 introduces delays in test execution due to long response times.
The cost of utilizing the API is another factor to consider, as it can be significant and needs to be weighed against the expected reduction in maintenance costs.
Future advancements and alternatives, such as locally installed LLMs, may address these concerns.
Overall, further research is necessary to fully understand the potential and limitations of using LLMs for web element identification.

\section{Future Work}\label{Future Work}

Even though the VON Similo LLM approach exceeds a 95\% success rate when locating the correct candidate, there are still almost 5\% to a perfect result.
Still, we do not know how the approach compares to humans since they might not reach 100\% success either.
However, we expect LLMs to become even more capable in the future.
They will also likely support more extensive prompt length (i.e., more tokens), become faster (i.e., lower response times), and the cost of using the APIs will decrease.

As a next step, we envision an approach that only relies on an LLM without needing a conventional algorithm to narrow down the number of candidates (e.g., VON Similo) that have the potential to enhance the effectiveness of web element localization further.
Such an approach could employ tournament selection~\cite{blickle1995mathematical} where all the visual web element candidates extracted from a web page attend, and the tournament winner is the selected candidate.
For example, assume 200 visual web elements extracted from a web page.
First, we divide the 200 candidates into ten groups of 20 candidates each.
The winner of each group will attend the final that selects the most similar candidate on the web page.
Our reasons for not trying such an approach today are: (1) a tournament would take a long time to complete since it requires many API requests, and (2) the cost would be high since the prompts will contain information gathered from all the web elements on the web page.
However, as advancements in LLM technology continue and API efficiency improves, the viability of such an approach may increase, making it promising for future exploration.

Another possible improvement is to provide the LLM with more information about the candidates to compare.
One such example could be a representation of the pictorial user interface (i.e., pixels) since that type of information is available to the human eye.
We decided to leave that out of our experiments since gathering and processing images from all visible images is likely time-consuming.
Also, there are many ways of processing and analyzing images, and exploring the alternatives would take lots of resources and time.

Instead of just asking the LLM once (i.e., one input returns one output) as in our experiment, we could employ other frameworks such as Chain of Thoughts (CoT) or Three Of Thoughts (ToT) that try to improve the results using a process of exploration of thoughts and self-evaluation \cite{yao2023tree}.
The drawback is, again, that more prompts increase the time and cost of using the API.

A possible way of increasing the efficiency and reducing the cost is to use VON Similo in cases when we expect it to be correct (i.e., a high probability) and only take advantage of the LLM in other cases.
This approach involves comparing the similarity score of the highest-ranked candidate with the remaining candidates to determine if it stands out as an outlier (i.e., clearly separated from the rest).
If a clear separation is detected, the top-ranked candidate from VON Similo is chosen as the result.
However, if no outlier is identified, the LLM is employed to decide among the top ten (or more) candidates.
This approach optimizes efficiency and cost by using the most appropriate model based on the probability of correctness and the distinctiveness of the top-ranked candidate.
The challenge with this approach is that imperfect detection of the outlier has a negative impact on the effectiveness since the LLM will not get the opportunity to find a better candidate.

The GPT API (all versions) is today provided as a cloud service.
One potential drawback of utilizing a cloud service is the inherent security risks associated with transmitting sensitive data to remote servers outside the company domain.
Relying on a third-party cloud provider might be a reason for not taking advantage of the benefits an LLM can provide regarding script robustness due to the possible security risk.
We might be able to solve this risk in the future by using an LLM that is powerful enough, and that can be locally installed, thus avoiding a cloud service.

\section{Acknowledgements}

This work was supported by the KKS foundation through the S.E.R.T. Research Profile project at Blekinge Institute of Technology. Robert Feldt has also been supported by the Swedish Scientific Council (No. 2015-04913, `Basing Software Testing on Information Theory').

\bibliography{mybibfile}

\begin{thebibliography}{10}
\providecommand{\url}[1]{#1}
\csname url@samestyle\endcsname
\providecommand{\newblock}{\relax}
\providecommand{\bibinfo}[2]{#2}
\providecommand{\BIBentrySTDinterwordspacing}{\spaceskip=0pt\relax}
\providecommand{\BIBentryALTinterwordstretchfactor}{4}
\providecommand{\BIBentryALTinterwordspacing}{\spaceskip=\fontdimen2\font plus
\BIBentryALTinterwordstretchfactor\fontdimen3\font minus
  \fontdimen4\font\relax}
\providecommand{\BIBforeignlanguage}[2]{{%
\expandafter\ifx\csname l@#1\endcsname\relax
\typeout{** WARNING: IEEEtran.bst: No hyphenation pattern has been}%
\typeout{** loaded for the language `#1'. Using the pattern for}%
\typeout{** the default language instead.}%
\else
\language=\csname l@#1\endcsname
\fi
#2}}
\providecommand{\BIBdecl}{\relax}
\BIBdecl

\bibitem{grechanik2009maintaining}
M.~Grechanik, Q.~Xie, and C.~Fu, ``Maintaining and evolving gui-directed test
  scripts,'' in \emph{Proceedings of the 31st international conference on
  software engineering}.\hskip 1em plus 0.5em minus 0.4em\relax IEEE Computer
  Society, 2009, pp. 408--418.

\bibitem{grechanik2009creating}
------, ``Creating gui testing tools using accessibility technologies,'' in
  \emph{Software Testing, Verification and Validation Workshops, 2009.
  ICSTW'09. International Conference on}.\hskip 1em plus 0.5em minus
  0.4em\relax IEEE, 2009, pp. 243--250.

\bibitem{olan2003unit}
M.~Olan, ``Unit testing: test early, test often,'' \emph{Journal of Computing
  Sciences in Colleges}, vol.~19, no.~2, pp. 319--328, 2003.

\bibitem{adamoli2011automated}
A.~Adamoli, D.~Zaparanuks, M.~Jovic, and M.~Hauswirth, ``Automated gui
  performance testing,'' \emph{Software Quality Journal}, vol.~19, no.~4, pp.
  801--839, 2011.

\bibitem{alegroth2013transitioning}
E.~Alegroth, R.~Feldt, and H.~H. Olsson, ``Transitioning manual system test
  suites to automated testing: An industrial case study,'' in \emph{2013 IEEE
  Sixth International Conference on Software Testing, Verification and
  Validation}.\hskip 1em plus 0.5em minus 0.4em\relax IEEE, 2013, pp. 56--65.

\bibitem{liebel2013state}
G.~Liebel, E.~Al{\'e}groth, and R.~Feldt, ``State-of-practice in gui-based
  system and acceptance testing: An industrial multiple-case study,'' in
  \emph{2013 39th Euromicro Conference on Software Engineering and Advanced
  Applications}, IEEE.\hskip 1em plus 0.5em minus 0.4em\relax -: -, 2013, pp.
  17--24.

\bibitem{mahmud2014design}
J.~Mahmud, A.~Cypher, E.~Haber, and T.~Lau, ``Design and industrial evaluation
  of a tool supporting semi-automated website testing,'' \emph{Software
  Testing, Verification and Reliability}, vol.~24, no.~1, pp. 61--82, 2014.

\bibitem{tonella2014recent}
P.~Tonella, F.~Ricca, and A.~Marchetto, ``Recent advances in web testing,'' in
  \emph{Advances in Computers}.\hskip 1em plus 0.5em minus 0.4em\relax
  Elsevier, 2014, vol.~93, pp. 1--51.

\bibitem{alegroth2017long}
E.~Al{\'e}groth and R.~Feldt, ``On the long-term use of visual gui testing in
  industrial practice: a case study,'' \emph{Empirical Software Engineering},
  vol.~22, no.~6, pp. 2937--2971, 2017.

\bibitem{dobslaw2019estimating}
F.~Dobslaw, R.~Feldt, D.~Micha{\"e}lsson, P.~Haar, F.~G. de~Oliveira~Neto, and
  R.~Torkar, ``Estimating return on investment for gui test automation
  frameworks,'' in \emph{2019 IEEE 30th International Symposium on Software
  Reliability Engineering (ISSRE)}, IEEE.\hskip 1em plus 0.5em minus
  0.4em\relax -: -, 2019, pp. 271--282.

\bibitem{nass2021many}
M.~Nass, E.~Al{\'e}groth, and R.~Feldt, ``Why many challenges with gui test
  automation (will) remain,'' \emph{Information and Software Technology}, vol.
  138, p. 106625, 2021.

\bibitem{choudhary2011water}
S.~R. Choudhary, D.~Zhao, H.~Versee, and A.~Orso, ``Water: Web application test
  repair,'' in \emph{Proceedings of the First International Workshop on
  End-to-End Test Script Engineering}, 2011, pp. 24--29.

\bibitem{montoto2011automated}
P.~Montoto, A.~Pan, J.~Raposo, F.~Bellas, and J.~L{\'o}pez, ``Automated
  browsing in ajax websites,'' \emph{Data \& Knowledge Engineering}, vol.~70,
  no.~3, pp. 269--283, 2011.

\bibitem{leotta2014reducing}
M.~Leotta, A.~Stocco, F.~Ricca, and P.~Tonella, ``Reducing web test cases aging
  by means of robust xpath locators,'' in \emph{2014 IEEE International
  Symposium on Software Reliability Engineering Workshops}.\hskip 1em plus
  0.5em minus 0.4em\relax IEEE, 2014, pp. 449--454.

\bibitem{leotta2016robula+}
------, ``Robula+: An algorithm for generating robust xpath locators for web
  testing,'' \emph{Journal of Software: Evolution and Process}, vol.~28, no.~3,
  pp. 177--204, 2016.

\bibitem{thummalapenta2012automating}
S.~Thummalapenta, S.~Sinha, N.~Singhania, and S.~Chandra, ``Automating test
  automation,'' in \emph{2012 34th International Conference on Software
  Engineering (ICSE)}.\hskip 1em plus 0.5em minus 0.4em\relax IEEE, 2012, pp.
  881--891.

\bibitem{leotta2015using}
M.~Leotta, A.~Stocco, F.~Ricca, and P.~Tonella, ``Using multi-locators to
  increase the robustness of web test cases,'' in \emph{Software Testing,
  Verification and Validation (ICST), 2015 IEEE 8th International Conference
  on}.\hskip 1em plus 0.5em minus 0.4em\relax IEEE, 2015, pp. 1--10.

\bibitem{nass2022similarity}
M.~Nass, E.~Al{\'e}groth, R.~Feldt, M.~Leotta, and F.~Ricca, ``Similarity-based
  web element localization for robust test automation,'' \emph{ACM Transactions
  on Software Engineering and Methodology}, 2022.

\bibitem{nass2023robust}
M.~Nass, R.~Coppola, E.~Al{\'e}groth, and R.~Feldt, ``Robust web element
  identification for evolving applications by considering visual overlaps,''
  \emph{arXiv preprint arXiv:2301.03863}, 2023.

\bibitem{vaswani2017attention}
A.~Vaswani, N.~Shazeer, N.~Parmar, J.~Uszkoreit, L.~Jones, A.~N. Gomez,
  {\L}.~Kaiser, and I.~Polosukhin, ``Attention is all you need,''
  \emph{Advances in neural information processing systems}, vol.~30, 2017.

\bibitem{feldt2023towards}
R.~Feldt, S.~Kang, J.~Yoon, and S.~Yoo, ``Towards autonomous testing agents via
  conversational large language models,'' \emph{arXiv preprint
  arXiv:2306.05152}, 2023.

\bibitem{kirinuki2022web}
H.~Kirinuki, S.~Matsumoto, Y.~Higo, and S.~Kusumoto, ``Web element
  identification by combining nlp and heuristic search for web testing,'' in
  \emph{2022 IEEE International Conference on Software Analysis, Evolution and
  Reengineering (SANER)}.\hskip 1em plus 0.5em minus 0.4em\relax IEEE, 2022,
  pp. 1055--1065.

\bibitem{liu2022fill}
Z.~Liu, C.~Chen, J.~Wang, X.~Che, Y.~Huang, J.~Hu, and Q.~Wang, ``Fill in the
  blank: Context-aware automated text input generation for mobile gui
  testing,'' \emph{arXiv preprint arXiv:2212.04732}, 2022.

\bibitem{liu2023chatting}
Z.~Liu, C.~Chen, J.~Wang, M.~Chen, B.~Wu, X.~Che, D.~Wang, and Q.~Wang,
  ``Chatting with gpt-3 for zero-shot human-like mobile automated gui
  testing,'' \emph{arXiv preprint arXiv:2305.09434}, 2023.

\bibitem{wu2023screen}
J.~Wu, A.~Swearngin, X.~Zhang, J.~Nichols, and J.~P. Bigham, ``Screen
  correspondence: Mapping interchangeable elements between uis,'' \emph{arXiv
  preprint arXiv:2301.08372}, 2023.

\bibitem{reppack}
\BIBentryALTinterwordspacing
(2023) Replication package. [Online]. Available:
  \url{https://github.com/michelnass/SimiloLLM}
\BIBentrySTDinterwordspacing

\bibitem{bubeck2023sparks}
S.~Bubeck, V.~Chandrasekaran, R.~Eldan, J.~Gehrke, E.~Horvitz, E.~Kamar,
  P.~Lee, Y.~T. Lee, Y.~Li, S.~Lundberg \emph{et~al.}, ``Sparks of artificial
  general intelligence: Early experiments with gpt-4,'' \emph{arXiv preprint
  arXiv:2303.12712}, 2023.

\bibitem{chowdhery2022palm}
A.~Chowdhery, S.~Narang, J.~Devlin, M.~Bosma, G.~Mishra, A.~Roberts, P.~Barham,
  H.~W. Chung, C.~Sutton, S.~Gehrmann \emph{et~al.}, ``Palm: Scaling language
  modeling with pathways,'' \emph{arXiv preprint arXiv:2204.02311}, 2022.

\bibitem{christiano2017deep}
P.~F. Christiano, J.~Leike, T.~Brown, M.~Martic, S.~Legg, and D.~Amodei, ``Deep
  reinforcement learning from human preferences,'' \emph{Advances in neural
  information processing systems}, vol.~30, 2017.

\bibitem{ouyang2022training}
L.~Ouyang, J.~Wu, X.~Jiang, D.~Almeida, C.~Wainwright, P.~Mishkin, C.~Zhang,
  S.~Agarwal, K.~Slama, A.~Ray \emph{et~al.}, ``Training language models to
  follow instructions with human feedback,'' \emph{Advances in Neural
  Information Processing Systems}, vol.~35, pp. 27\,730--27\,744, 2022.

\bibitem{wood1998document}
L.~Wood, A.~Le~Hors, V.~Apparao, S.~Byrne, M.~Champion, S.~Isaacs, I.~Jacobs,
  G.~Nicol, J.~Robie, R.~Sutor \emph{et~al.}, ``Document object model (dom)
  level 1 specification,'' \emph{W3C recommendation}, vol.~1, 1998.

\bibitem{nass2023similarity}
M.~Nass, E.~Al{\'e}groth, R.~Feldt, M.~Leotta, and F.~Ricca, ``Similarity-based
  web element localization for robust test automation,'' \emph{ACM Transactions
  on Software Engineering and Methodology}, vol.~32, no.~3, pp. 1--30, 2023.

\bibitem{10132199}
M.~Nass, E.~Alégroth, R.~Feldt, and R.~Coppola, ``Robust web element
  identification for evolving applications by considering visual overlaps,'' in
  \emph{2023 IEEE Conference on Software Testing, Verification and Validation
  (ICST)}, 2023, pp. 258--268.

\bibitem{koubaa2023gpt}
A.~Koubaa, ``Gpt-4 vs. gpt-3.5: A concise showdown,'' 2023.

\bibitem{alegroth2016maintenance}
E.~Al{\'e}groth, R.~Feldt, and P.~Kolstr{\"o}m, ``Maintenance of automated test
  suites in industry: An empirical study on visual gui testing,''
  \emph{Information and Software Technology}, vol.~73, pp. 66--80, 2016.

\bibitem{jain1996artificial}
A.~K. Jain, J.~Mao, and K.~M. Mohiuddin, ``Artificial neural networks: A
  tutorial,'' \emph{Computer}, vol.~29, no.~3, pp. 31--44, 1996.

\bibitem{y2022large}
B.~A. y~Arcas, ``Do large language models understand us?'' \emph{Daedalus},
  vol. 151, no.~2, pp. 183--197, 2022.

\bibitem{min2022rethinking}
S.~Min, X.~Lyu, A.~Holtzman, M.~Artetxe, M.~Lewis, H.~Hajishirzi, and
  L.~Zettlemoyer, ``Rethinking the role of demonstrations: What makes
  in-context learning work?'' \emph{arXiv preprint arXiv:2202.12837}, 2022.

\bibitem{razeghi2022impact}
Y.~Razeghi, R.~L. Logan~IV, M.~Gardner, and S.~Singh, ``Impact of pretraining
  term frequencies on few-shot reasoning,'' \emph{arXiv preprint
  arXiv:2202.07206}, 2022.

\bibitem{10.1145/2950290.2950294}
\BIBentryALTinterwordspacing
M.~Hammoudi, G.~Rothermel, and A.~Stocco, ``Waterfall: An incremental approach
  for repairing record-replay tests of web applications,'' in \emph{Proceedings
  of the 2016 24th ACM SIGSOFT International Symposium on Foundations of
  Software Engineering}, ser. FSE 2016.\hskip 1em plus 0.5em minus 0.4em\relax
  New York, NY, USA: Association for Computing Machinery, 2016, p. 751–762.
  [Online]. Available: \url{https://doi.org/10.1145/2950290.2950294}
\BIBentrySTDinterwordspacing

\bibitem{Kirinuki2019COLORCL}
H.~Kirinuki, H.~Tanno, and K.~Natsukawa, ``Color: Correct locator recommender
  for broken test scripts using various clues in web application,'' \emph{2019
  IEEE 26th International Conference on Software Analysis, Evolution and
  Reengineering (SANER)}, vol.~36, no.~4, pp. 310--320, 2019.

\bibitem{khaliq2022transformers}
Z.~Khaliq, S.~U. Farooq, and D.~A. Khan, ``Transformers for gui testing: A
  plausible solution to automated test case generation and flaky tests,''
  \emph{Computer}, vol.~55, no.~3, pp. 64--73, 2022.

\bibitem{levenshtein}
\BIBentryALTinterwordspacing
``Levenshtein.'' [Online]. Available: \url{\url{http://levenshtein.net}}
\BIBentrySTDinterwordspacing

\bibitem{brisset2022erratum}
S.~Brisset, R.~Rouvoy, L.~Seinturier, and R.~Pawlak, ``Erratum: Leveraging
  flexible tree matching to repair broken locators in web automation scripts,''
  \emph{Information and Software Technology}, vol. 144, p. 106754, 2022.

\bibitem{grechanik2018differencing}
M.~Grechanik, C.~W. Mao, A.~Baisal, D.~Rosenblum, and B.~M. Hossain,
  ``Differencing graphical user interfaces,'' in \emph{2018 IEEE International
  Conference on Software Quality, Reliability and Security (QRS)}.\hskip 1em
  plus 0.5em minus 0.4em\relax IEEE, 2018, pp. 203--214.

\bibitem{xie2009guide}
Q.~Xie, M.~Grechanik, C.~Fu, and C.~Cumby, ``Guide: A gui differentiator,'' in
  \emph{2009 IEEE International Conference on Software Maintenance}.\hskip 1em
  plus 0.5em minus 0.4em\relax IEEE, 2009, pp. 395--396.

\bibitem{liu2022crawlabel}
Y.~Liu, R.~Yandrapally, A.~K. Kalia, S.~Sinha, R.~Tzoref-Brill, and A.~Mesbah,
  ``Crawlabel: computing natural-language labels for ui test cases,'' in
  \emph{Proceedings of the 3rd ACM/IEEE International Conference on Automation
  of Software Test}, 2022, pp. 103--114.

\bibitem{yandrapally2014robust}
R.~Yandrapally, S.~Thummalapenta, S.~Sinha, and S.~Chandra, ``Robust test
  automation using contextual clues,'' in \emph{Proceedings of the 2014
  International Symposium on Software Testing and Analysis}, 2014, pp.
  304--314.

\bibitem{nguyen2021generating}
V.~Nguyen, T.~To, and G.-H. Diep, ``Generating and selecting resilient and
  maintainable locators for web automated testing,'' \emph{Software Testing,
  Verification and Reliability}, vol.~31, no.~3, p. e1760, 2021.

\bibitem{leotta2021sidereal}
M.~Leotta, F.~Ricca, and P.~Tonella, ``Sidereal: Statistical adaptive
  generation of robust locators for web testing,'' \emph{Software Testing,
  Verification and Reliability}, vol.~31, no.~3, p. e1767, 2021.

\bibitem{blickle1995mathematical}
T.~Blickle and L.~Thiele, ``A mathematical analysis of tournament selection.''
  in \emph{ICGA}, vol.~95.\hskip 1em plus 0.5em minus 0.4em\relax Citeseer,
  1995, pp. 9--15.

\bibitem{yao2023tree}
S.~Yao, D.~Yu, J.~Zhao, I.~Shafran, T.~L. Griffiths, Y.~Cao, and K.~Narasimhan,
  ``Tree of thoughts: Deliberate problem solving with large language models,''
  \emph{arXiv preprint arXiv:2305.10601}, 2023.

\end{thebibliography}

\end{document}